\documentclass[letters,twocolumn,floatix,amssymb,showpacs,amsmath, superscriptaddress,longbibliography]{revtex4-1}
\usepackage{graphicx}
\usepackage{upgreek}
\usepackage{epsfig,color}
\usepackage{amsmath,amssymb,amsthm}
\usepackage[utf8]{inputenc}
\usepackage[colorlinks, urlcolor=blue, linkcolor=red]{hyperref}

\usepackage{soul}
\usepackage{ dsfont }
\usepackage{tikz}
\usepackage[normalem]{ulem}

\newcommand{\beqa}{\begin{eqnarray}}
\newcommand{\eeqa}{\end{eqnarray}}

\renewcommand{\boxed}[2]{\textcolor{#1}{%
\tikz[baseline={([yshift=-1ex]current bounding box.center)}] \node [rectangle, minimum width=1ex,rounded corners,draw] {\normalcolor\m@th$\displaystyle#2$};}}

\usepackage{bookmark}

\usepackage{pgfplots}
\usepackage{pgfplotstable}

\pgfplotstableread[col sep=comma]{QLnodetun.csv}{\QLnodetun}
\pgfplotstableread[col sep=comma]{QLdetun2.csv}{\QLdetunpp}
\pgfplotstableread[col sep=comma]{QLdetun3.csv}{\QLdetunppp}
\pgfplotstableread[col sep=comma]{CLHetphNum.csv}{\CLHetphNum}
\pgfplotstableread[col sep=comma]{CLULphNum.csv}{\CLULphNum}
\pgfplotstableread[col sep=comma]{OptSqData1.csv}{\OptSqDatap}
\pgfplotstableread[col sep=comma]{CLhetloss.csv}{\CLhetloss}
 \pgfplotstableread[col sep=comma]{CLULloss.csv}{\CLULloss}

\begin{document}

\title{Heisenberg-Limited Quantum Lidar for Joint Range and Velocity Estimation}
\author{Maximilian Reichert}
\email{maximilian.reichert@ehu.eus}
\affiliation{Department of Physical Chemistry, University of the Basque Country UPV/EHU, Apartado 644, 48080 Bilbao, Spain}
\affiliation{EHU Quantum Center, University of the Basque Country UPV/EHU, Apartado 644, 48080 Bilbao, Spain}

\author{Quntao Zhuang}
\affiliation{Ming Hsieh Department of Electrical and Computer Engineering,
University of Southern California, Los Angeles, CA 90089, USA}
\affiliation{Department of Physics and Astronomy, University of Southern California, Los Angeles, CA 90089, USA}

\author{Mikel Sanz}
\affiliation{Department of Physical Chemistry, University of the Basque Country UPV/EHU, Apartado 644, 48080 Bilbao, Spain}
\affiliation{EHU Quantum Center, University of the Basque Country UPV/EHU, Apartado 644, 48080 Bilbao, Spain}
\affiliation{IKERBASQUE, Basque Foundation for Science, Plaza Euskadi 5, 48009, Bilbao, Spain}
\affiliation{Basque Center for Applied Mathematics (BCAM), Alameda de Mazarredo, 14, 48009 Bilbao, Spain}

\date{\today}

\begin{abstract}
We propose a quantum lidar protocol to jointly estimate the range and velocity of a target by illuminating it with a single beam of pulsed displaced squeezed light. In the lossless scenario, we show that the mean-squared errors of both range and velocity estimations are inversely proportional to the squared number of signal photons, simultaneously attaining the Heisenberg limit. This is achieved by engineering the multi-photon squeezed state of the temporal modes and adopting standard homodyne detection. To assess the robustness of the quantum protocol, we incorporate photon losses and detuning of the homodyne receiver. Our findings reveal a quantum advantage over the best-known classical strategy across a wide range of round-trip transmissivities. Particularly, the quantum advantage is substantial for sufficiently small losses, even when compared to the optimal---potentially unattainable---classical performance limit. The quantum advantage also extends to the practical case where quantum engineering is done on top of a strong classical coherent state with watts of power. This, together with the robustness against losses and the feasibility of the measurement with state-of-the-art technology, make the protocol highly promising for near-term implementation.

\end{abstract}

\maketitle

A classical lidar (CL) estimates multiple parameters of a remote target by illuminating it with a single beam of coherent light and measuring the reflected signal. The signal flight time is used to estimate the range, while measuring the frequency shift due to the Doppler effect is used to estimate the velocity. The classical ultimate accuracy limit,
i.e. the Cram\'er-Rao bound is a lower bound of the estimate's mean-squared error (MSE), and scales inversely with the number of signal photons $\sim N^{-1}$ for both range and velocity \cite{helstrom}. 

A quantum lidar (QL) works similarly to the CL, but employs quantum effects to enhance the performance. Thus far,  the proposed protocols in Refs.~\cite{reichert2022,giovannetti2001,giovannetti2002,shapiro2007,maccone2020,maccone2023} all rely on  entanglement to reach the Heisenberg limit (HL), i.e. a scaling of the MSE $\sim N^{-2}$, for at least one of the parameters. The proposals of Refs.~\cite{zhuang2017,huang2021,li2023} considered the task of simultaneous range and velocity estimation. These protocols achieve a quantum enhancement by utilizing a pair of entangled signal and idler beams which, however, complicates experimental implementation. So far, only in Ref.~\cite{zhuang2017} the simultaneous attainability of the HL for both parameters was demonstrated. However, the employed state is at the single photon level and cannot tolerate the loss of even a single photon, and the required measurement therein is based on unfeasible unit efficiency sum frequency generation.

In this Letter, we propose a quantum lidar protocol that simultaneously achieves the Heisenberg limit for estimating the range and velocity of a target. This protocol mimics the classical lidar setup~\cite{helstrom}, employing a single spatial mode and homodyne detection, facilitating the comparison between the classical and quantum protocols. The quantum probe pulse consists of modes in displaced squeezed states, which can be experimentally produced via spontaneous parametric down conversion (SPDC) pumped by a laser pulse~\cite{wasilewski,roman2023}. In contrast to previous proposals, this protocol uses squeezing, and does not require an entangled idler beam, i.e. a quantum memory. 
In addition, we consider the major imperfections in the optical domain---photon losses and detunings in the homodyne receiver---and demonstrate the protocol's resilience.
By optimizing the displacement in different operating conditions, we demonstrate a significant quantum advantage against the best-known classical strategy for most values of the round-trip transmissivities. The quantum advantage also extends to the practical case where quantum engineering is done on top of the strong classical coherent state with watts of power. This is a notable advancement considering that previous quantum-enhanced lidar protocols~\cite{giovannetti2001,giovannetti2002,shapiro2007, maccone2020,maccone2023,zhuang2017,huang2021} use experimentally cumbersome states and/or measurements, and lose quantum advantage after the loss of a single photon.

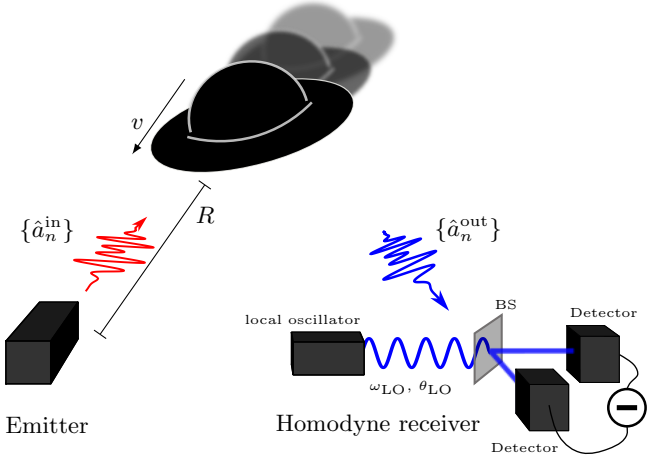
\begin{figure}
\centering
\begin{tikzpicture}
    \node[anchor=south west,inner sep=-1] (image) at (0,0) {\includegraphics[width=\columnwidth]{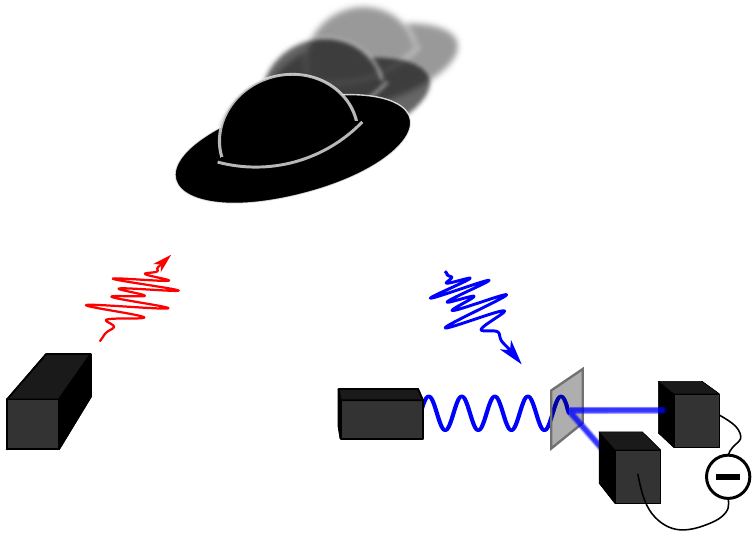}};
    \node at (4,1.8) {\tiny local oscillator};
    \node at (5.4,0.9) {\tiny{ $ \omega_{\text{LO}}$, $\theta_{\text{LO}}$ }};
     \node at (0.6,0.4) {Emitter};
     \node at (0.6,3) {$\{\hat{a}_n^{\text{in}}\}$};
       \node at (6.2,3) {$\{\hat{a}_n^{\text{out}}\}$};
     \node at (5,0.4) {Homodyne receiver};
     \node at (6.7,2) {\tiny BS};
     \node at (8,1.9) {\tiny Detector};
     \node at (6.95,0.1) {\tiny Detector};
        \draw[|-|] (1.3,1.6) -- (2.7,3.6);
        \node at (2.7,3.2)  {$R$};
         \draw[latex-] (1.73,4) -- (2.43,5);
        \node at (1.8,4.4)  {$v$};
\end{tikzpicture}
\caption{Quantum lidar protocol. The signal beam is sent towards the target located at range $R$ with velocity $v$. The homodyne receiver mixes the reflected signal via a balanced beam splitter (BS)  with a continuous wave local oscillator with frequency $\omega_{\text{LO}}$ and phase $\theta_{\text{LO}}$. The two outputs are then measured by photo detectors and the resulting photo currents are subtracted from each other, giving rise to the homodyne current. } \label{fig:model}
\end{figure}

{\em Quantum model of range and velocity estimation.---} 
Let us now introduce the model used for our QL and the CL benchmark depicted in Fig~\ref{fig:model}. We adopt the formalism developed in Ref.~\cite{shapiro2009}. We consider a single (paraxial) $+z$ propagating spatial mode of the electromagnetic field (EM) in the $xy$ plane at the location $z=0$ of the emitter. Assuming  the signal to be quasi monochromatic, we can approximate the positive frequency part of the EM field operator in the Heisenberg picture as
\begin{equation}
    \hat{E}(t) = \frac{1}{\sqrt{2\pi}} \int_{-\infty}^\infty\mathrm{d}  \nu  \,  \hat{A} (\nu) e^{-i   \nu   t} ,
\end{equation}
with commutator relations $[\hat{E}(t), \hat{E}^\dag(t')]= \delta (t-t')$ and $[\hat{A} (\nu),\hat{A}^\dag(\nu')] = \delta (\nu-\nu')$. The operator $\hat{E}^\dag (t)$ creates a photon at time $t$ in the detector plane, while $\hat{A}^\dag(\nu)$ creates an unlocalized photon of frequency $ \nu $. The lidar will emit a signal of time duration $\Delta T$, bandwidth $\Delta\omega$ (definition in Supplemental Material (SM)~\ref{app:sq}) and average photon number $N = \int \mathrm{d} t \langle \hat{E}^\dag(t) \hat{E}(t)\rangle$.

Let us now come to the model of the target. For the ideal case, we assume diffraction to be negligible and the target to be perfectly reflecting mirror in the $xy$ plane located at radial distance $R$ with radial velocity $v$ at the time $\tau_0+ R/c$ of signal-target interaction. As has been shown in Refs.~\cite{reichert2022,zhuang2017,good2013}, the interaction between probe state $\vert \psi^{\text{in}} \rangle $ and target induces in the non-relativistic limit $v/c \ll 1$, where $c$ is the speed of light, approximately the transformation  $\hat{E} (t) \rightarrow \hat{E} (t+ 2 R/c) e^{ i ( 2\omega_0 v t/c +\theta'_R)}$ on the state acting operators (Schr\"odinger picture), where $\theta_R'$ is some unknown phase shift. We also note that the propagation direction changes to $-z$ going and that we neglect the small deformation of the bandwidth parameter~\cite{reichert2022,VanTreesIII}. In addition, we model the round-trip loss, that accounts for imperfect reflection, diffraction and absorption, as a beam splitter of transmissivity  $\kappa$ applied before measurement. 

Motivated by experimental considerations of SPDC pumped with a laser pulse \cite{wasilewski,roman2023,presutti} and algebraic convenience, we utilize the set of orthogonal Hermite-Gaussian (HG) modes as the basis in which we describe our state, which are given by
\begin{multline}\label{eq:modes}
    \phi_n^{\text{in}} (t) = \Phi_n(t;\tau_0,\omega_0,\theta_0,\sigma)\\\equiv \frac{\sqrt{\sigma} \sqrt{\frac{2^{-n}}{n!}}}{\sqrt[4]{2 \pi }}   H_n\left(\frac{\sigma
   (t-\tau_0 )}{\sqrt{2}}\right) e^{ -\frac{1}{4} \sigma
   ^2 (t-\tau_0 )^2} e^{-i (\omega_0 t + \theta_0)} ,
\end{multline}
where $H_n (t) = (-1)^n e^{t^2} \frac{\mathrm{d}^n}{\mathrm{d}t^n} e^{-t^2}$ are the physicist's Hermite polynomials. The mode functions are centered around time $\tau_0 $, with carrier frequency $\omega_0$ and bandwidth parameter $\sigma$, for more details see SM~\ref{app:Modes}. We can now define mode operators
$
    \hat{a}_n^{\text{in}} = \int_{-\infty}^\infty   \mathrm{d}t \, \phi_n^{\text{in}*} (t) \hat{E}(t) 
$.
The class of states we will consider are Gaussian states, i.e.  displaced squeezed states of the form
\begin{align} \label{eq:instate}
   \vert \psi^{\text{in}}\rangle = \bigotimes_{n=0}^{\infty} [\hat{D}_n^\text{in} (\alpha_n) \hat{S}_n^\text{in} (\zeta_n)]\vert 0\rangle
\end{align}
where $\alpha_n ,\zeta_n \in \mathds{C}$ and $\zeta_n = r_n e^{i\varphi_n}$ with $r_n >0$. For more information about these states see the SM~\ref{app:sq}.
Later, we will discuss how to engineer the mode configuration in different operating regimes to obtain quantum advantage.


Due to the interaction with the target, the mode operators transform as $\hat{a}_n^{\text{in}} \rightarrow   \hat{a}_n^{\text{out}} = \int_{-\infty}^\infty   \mathrm{d}t \, \phi_n^{\text{in}*} (t)  \hat{E} (t+ 2 R/c) e^{ i ( 2\omega_0 v t/c +\theta_R')}$. A change of variables, i.e. $ \hat{a}_n^{\text{out}} = \int_{-\infty}^\infty   \mathrm{d}t \, \phi_n^{\text{out}*} (t) \hat{E}(t)$, reveals that the mode parameters change according to  
\begin{subequations}
\label{eq:parameter_change}
\begin{align}
    \tau_0 &\rightarrow \tau = \tau_0 + 2R/c  \label{eq:delay}\\
    \omega_0 &\rightarrow \omega =  \omega_0 + 2\omega_0 v /c \label{eq:frequency} \\ 
    \theta_0 &\rightarrow \theta = \theta_0 + \theta_R  , \label{eq:phase}
\end{align}
\end{subequations}
such that $\phi_n^{\text{out}} (t) =\Phi_n(t;\tau,\omega,\theta,\sigma)$ takes the identical form as in Eq.~\eqref{eq:modes} with updated parameters from above. As standard in the  literature~\cite{helstrom, zhuang2017, zhuang2022}, we ignore the information about $R$ and $v$ encoded in the phase shift $\theta_R = \theta_R (\theta_R',R,v)$ due to phase ambiguity and fluctuation. The reflected state is given by 
$\vert \psi^{\text{out}}\rangle = \bigotimes_{n=0}^{\infty} [\hat{D}_n^\text{out} (\alpha_n) \hat{S}_n^\text{out} (\zeta_n)]\vert 0\rangle$, taking the same form as the previously defined $\vert \psi^{\text{in}}\rangle$, with the only modification that the in operators are replaced with out operators while $\{\alpha_n \}_n$ and $\{\zeta_n \}_n$ remain unchanged.
  From Eqs.~\eqref{eq:modes},\eqref{eq:delay}-\eqref{eq:phase}, we can solve the induced mode transform for infinitesimal small parameter changes $\epsilon_\tau$, $\epsilon_\omega$ and $\epsilon_\theta$ (see SM~\ref{app:Modes}),
\begin{align} 
\hat{a}_n^{\text{out}} \approx \hat{a}_n^\text{in} A_n + \hat{a}_{n-1}^\text{in} B_n -\hat{a}_{n+1}^\text{in} B_{n+1}^\star
\label{mode_transform}
\end{align}
for $n\ge 1$ with $A_n = 1+ i \epsilon_\theta + i \epsilon_\omega \tau_0$  and $B_n = -\epsilon_\tau \frac{\sigma\sqrt{n}}{2}+i \epsilon_\omega \frac{\sqrt{n}}{\sigma}$;
in the $n=0$ case, we can ignore the $\hat{a}_{-1}$ term above. From above, we see  in terms of the temporal modes, that the transformation induces a phase shift $i \epsilon_\omega \tau_0$ and mode mixing between neighbors with coefficients almost conjugate to each other. Analogous to Ref.~\cite{xia2020} where frequency modes are mixed, mode mixing of Eq.~\eqref{mode_transform} indicates that displacement and squeezing are both needed to achieve the best performance. 
Equation~\eqref{mode_transform} seems to indicate that higher-order time-modes with large $n$ will enhance the precision, as $|B_n|$ is large; which is indeed true, however at the cost of increasing the time-bandwidth product. We will show that the quantum protocol increases the precision without increasing the time-bandwidth.

The next step of the protocol is measuring the reflected state. We will consider homodyne and heterodyne reception, which was recently considered as well in quantum radar protocols \cite{reichert2023,Shi2022,Shi2023}. In homo/heterodyne detection, the returned signal is  mixed with a strong local oscillator, i.e. a coherent state with temporal mode $\sim e^{-i (\omega_{\text{LO}}t + \theta_{\text{LO}})}$, where $\omega_{\text{LO}}$ is the local oscillator (LO) frequency and $\theta_{\text{LO}}$ is the LO  phase.  This measurement is represented by the operator~\cite{Lvovsky,qin2015} (see also SM~\ref{app:homodyne})
\begin{equation} \label{eq:homodyneoperator}
    \hat{X} (t) = \frac{\hat{E}(t) e^{i\omega_{\text{LO}}t +i\theta_{\text{LO}}} + \hat{E}^\dag (t) e^{-i\omega_{\text{LO}}t-i\theta_{\text{LO}}} }{\sqrt{2}} ,
\end{equation}
whose measurement outcome $X(t)$ is a continuous function of time. This measurement is Gaussian, and for Gaussian states like ours, the statistics of $X(t)$ are Gaussian distributed \cite{serafini}. The corresponding mean $\mu (t) = \langle \hat{X}(t) \rangle$ and covariance function $\Sigma (t,t') = (\langle \hat{X}(t)-\mu (t)) (\hat{X}(t') -\mu (t'))\rangle $ can be calculated using the out mode decomposition  $\hat{E}(t) = \sum_n \phi_n^{\text{out}}(t) \hat{a}_n^{\text{out}}$. We find for our general displaced squeezed out state $\vert\psi^{\text{out}}\rangle$ the mean $\mu (t) = \sqrt{2\kappa}  \text{Re}[ \sum_n \alpha_n \Bar{\phi}_n^{\text{out}}(t) e^{i (\delta \omega t+\delta \theta)}]$ and covariance function  (see SM~\ref{app:sqFIM})
\begin{widetext}
\begin{equation}
  \Sigma (t,t')  = \frac{1}{2} \delta (t-t') + \kappa \sum_{n=0}^\infty \bar{\phi}_n^{\text{out}} (t)\bar{\phi}_n^{\text{out}} (t') \left[ 
N_{\text{sq},n} \cos \left( \delta\omega \, (t-t') \right) - \sqrt{N_{\text{sq},n} (N_{\text{sq},n}+1)} \cos \left( \delta\omega \, (t+t')+2\delta\theta+\varphi_n \right) \right] ,
 \label{eq:covariance}
\end{equation}
\end{widetext}
where $N_{\text{sq},n} = \sinh^2 ( r_n )$ and we defined the detunings $\delta\omega = \omega_{\text{LO}}-\omega$ and $\delta\theta = \theta_{\text{LO}}-\theta$,  and we define $\bar{\phi}_n^{\text{out}}(t)$  as the envelope of the (out) mode $\phi_n^{\text{out}} (t) = \bar{\phi}_n^{\text{out}}(t) e^{-i(\omega t +\theta)}$. In the case of perfect signal matching $\delta\omega = \delta\theta=0$, measuring $\hat{X}(t)$ for all times $t$ is equivalent to measuring the quadratures $\{(\hat{a}_n^{\text{out}}+\hat{a}_n^{\text{out}\dag})/\sqrt{2} \}_n$ for all $n$ (see Ref.~\ref{app:homodyne}). If $\delta\omega \Delta T \ll 1$, we call the measurement homodyne detection of phase $\theta_{\text{LO}}$, otherwise we call it heterodyne detection.
Here, we analyze arbitrary detunings and a finite time resolution $\Delta t$ small enough to resolve the mean and covariance function. We obtain the discretized operators $\hat{X}_i =  \int_{i\Delta t}^{(i+1)\Delta t} \mathrm{d}t \, \hat{X}(t)/\Delta t$, whose outcomes are Gaussian distributed with mean $\mu_i= \langle  \hat{X}_i \rangle$ and covariance matrix $\Sigma_{ij} = \langle  (\hat{X}_i - \langle \hat{X}_i\rangle)  (\hat{X}_j - \langle \hat{X}_j\rangle) \rangle $~\cite{serafini}.

We will consider the estimation of the parameters $\tau$ and $\omega$ instead  of $R$ and $v$ as they appear more naturally in our model. The phase $\theta$ will be treated as a nuisance parameter.  For the estimation task, we use estimators $\tau_{\text{est}}(\{ X_i \}_i)$ and $\omega_{\text{est}} (\{ X_i \}_i)$ that map the observed measurement data to estimates. To evaluate the ultimate performance that can be achieved with optimal unbiased estimators, we use the Fisher information matrix (FIM) $F$, whose components for Gaussian measurement distributions are given by~\cite{VanTrees2001}
$
    F_{\alpha \beta} = \partial_{\alpha} \vec{\mu} \cdot \Sigma^{-1} \cdot \partial_{\beta} \vec{\mu} +  \frac{1}{2} \text{Tr}\left[ \Sigma^{-1}\cdot \partial_\alpha \Sigma \cdot\Sigma^{-1} \cdot \partial_\beta \Sigma \right]
$,
where $\alpha,\beta \in \{\tau,\omega,\theta \}$.
The diagonal elements of the inverse FIM $F^{-1}_{\hphantom{-1}\alpha\alpha}$  set the ultimate accuracy limits via the Cram\'er-Rao bound (CRB). The CRB is a lower bound of the variance (i.e. MSE) of unbiased parameter estimators~\cite{VanTrees2001}
\begin{align}\label{eq:CRB}
    \text{Var} [\alpha_{\text{est}}] \geq \frac{F^{-1}_{\hphantom{-1}\alpha\alpha}}{M} ,
\end{align}
where $M$ is the number of identical experiment repetitions. The lower bound can be achieved for $M=1$ if good prior information of parameters is available~\cite{rafal2020,meyer2023}, or in general when repetition $M\gg1$.
In the case of a lidar, classical or quantum, many identical repetitions can be realized by sending a train of identical pulses towards the target assuming its state does not change much during the total duration of the pulse train.

So far, we have considered a fixed measurement. By considering the optimal measurement for each parameter, one obtains the quantum Fisher information matrix (QFIM) $J$, which for mode parameters and general states can be calculated using formulas given in Ref.~\cite{gessner2023,sorelli} and in SM~\ref{app:QFIMmode}. The QFIM for our state with parameters $\{\tau,\omega,\theta \}$ is given in the SM~\ref{app:sqFIM}. The quantum Cram\'er-Rao bound is given by $  \text{Var} [\alpha_{\text{est}}] \geq J^{-1}_{\hphantom{-1}\alpha\alpha}/M$. For the multiparameter case, it is generally not attainable simultaneously for all parameters, due to the fact that the optimal measurements for each parameter do not commute~\cite{liu}.

{\em Coherent state lidar.---}
Let us first consider the classical benchmark protocol that will later be compared with the quantum lidar protocol. The classical benchmark employs a single beam coherent state, which is a special case of Eq.~\eqref{eq:instate} with no squeezing, i.e. $\zeta_n =0$ for all $n$. For the sake of obtaining compact expressions, we only consider received signals that are not frequency modulated. As we show in the SM~\ref{app:cohQFIM}, frequency modulation does not yield a benefit in joint estimation. We obtain analytically the inverse QFIM elements linked to the QCRBs (see SM~\ref{app:cohQFIM})
$
    J^{-1}_{\text{C} \,\,\,\tau\tau} = 1/4 \Delta \omega^2 N$ and $ J^{-1}_{\text{C} \,\,\,\omega \omega} =1/4 \Delta T^2  N,
$
which are the classical ultimate accuracy limits.
Note, that the QCRBs do not depend on the specific shape of the signal, rather they only depend on the time duration $\Delta T$ and bandwidth $\Delta\omega$. As we show in the SM~\ref{app:cohFIM}, the homodyne measurement can be optimal for one of the parameters, but not optimal for both simultaneously. 
We are not aware of any measurement that achieves this optimal precision for both parameters simultaneously.
Instead, we can benchmark with the best known CL protocol that employs heterodyne detection. Utilizing the Gaussian nature of the statistics, we can evaluate the CRB analytically,
\begin{align}
    F^{-1}_{\text{C,het} \, \tau \tau} = \frac{1}{2 \Delta \omega^2 N}, \quad F^{-1}_{\text{C,het} \,\omega\omega} =\frac{1}{ 2 \Delta T^2   N } ,
\end{align}
which is only a prefactor worse than the ultimate (potentially unachievable) classical performance limit and recovers the results from classical literature~\cite{helstrom}.

{\em Displaced squeezed state quantum lidar.---}
Let us now consider our QL protocol with the general displaced squeezed state given in Eq.~\ref{eq:instate} with displacement $\{ \alpha_n\}_n$ and squeezing $\{ \zeta_n\}_n$ parameters. For this general state we calculated in Ref.~\ref{app:sqFIM} the FIM of $\hat{X}(t)$ for parameters $\tau$, $\omega$ and $\theta$, using the mean and covariance function given in Eq.~\eqref{eq:covariance}. We assumed $\delta\omega \Delta T\ll 1$ (homodyne condition), arbitrary $\delta\theta$ and arbitrary path transmissivity $\kappa$. 

There is a huge landscape of possible parameter configurations. As indicated by Eq.~\eqref{mode_transform}, it is important to have at least two modes occupied to make use of the mode mixing. Also, we see that the squeezing angles need to be adjusted between $0$ and $\pi/2$ to enhance the estimation of both real and imaginary part so that multiple parameters can be extracted.
Here, we will examine the mode configuration $\hat{D}_1 (\sqrt{N_{\text{coh}}}e^{-i\pi/4}) \hat{S}_{0}(r)\hat{S}_1 (r e^{i\pi/2}) \hat{S}_2 (r) \vert 0 \rangle$. Let us first consider the ideal scenario $\kappa =1$. In the case of $N\gg 1$, we obtain a compact expression for the inverse FIM diagonal, yielding the CRBs
\begin{align}\label{eq:CRBs}
      F^{-1}_{\hphantom{-1}\tau \tau} = \frac{1}{ \Delta \omega^2 N^2}, \quad F^{-1}_{\hphantom{-1}\omega\omega} =\frac{1}{  \Delta T^2   N^2 } .
\end{align}
In the SM~\ref{app:sqQFIM} we also calculated the corresponding  QCRBs and found that they are equal to the CRBs, implying that the QLs optimal measurement is homodyne detection.
As discussed earlier, unlike the QCRB, the CRB is always simultaneously attainable both in a local sense \cite{rafal2020} and in an asymptotic sense of many repetitions. We thus have proven that the Heisenberg limit is jointly achievable for range and velocity of a target which is our main result. 
In Eq.~\eqref{eq:CRBs}, we optimized the energy's allocation  into displacement and squeezing  $N= N_{\text{coh}}+N_{\text{sq}}$  with $N_{\text{sq}} = 3N/4$ and $N_{\text{coh}} = N/4$. Other distributions of squeezing and displacement (with $N_{\text{sq}}\neq 0$) also reach the HL with a  worse prefactor. Also a fully squeezed state with $N_{\text{coh}}=0$ reaches the HL.  Estimators that reach the joint HL were found and given in SM~\ref{app:estimators}. 
To achieve the joint HL, the LO's frequency and phase have to be matched to the returned signal such that $\vert\delta\omega \Delta T \vert \ll 1$ and  $\vert\delta\theta +\delta\omega\tau \vert \ll 1/(N_{\text{sq}}+1)$ (SM~\ref{app:estimators}). This implies sufficiently good prior knowledge of the parameters $\tau$, $\omega$ and $\theta$, which may be obtained from a second prior-acquiring classical pulse that could be sent in a different frequency band. Also note, that  other measurements exist that do not require priors, namely time-resolved and frequency-resolved photon counting, which are optimal for single parameter range and velocity estimation respectively Ref.\cite{reichert2022}. However, at least two shots are necessary to extract both parameters with these alternative measurements. When the energy is equally distributed between these two shots, the variance of each parameter worsens by a factor of 4 due to $1/(N/2)^2= 4/N^2$, which is why a single shot strategy is preferable.¡

\begin{figure*}[!] 
\begin{tikzpicture} 
    \begin{axis}[
    scale only axis, 
      width=0.8\columnwidth, 
      height=0.35\columnwidth, 
    ymode=log,
    xmode=log,
    xmin = 0.1,
    xmax=10000,
    xlabel={photon number $N$},
    ylabel={$\text{Var}[\tau]\cdot\text{Var}[\omega]$},
    legend style={font=\tiny, row sep=-0.0cm},
    legend pos=south west, 
     legend cell align={left}, 
    ]
  ]
    \addplot[blue, thick, smooth] table {\QLnodetun};
    \addlegendentry{QL} 
              \addplot[blue, loosely dotted, thick, smooth] table {\QLdetunpp};
       \addlegendentry{QL $\delta\theta = 0.01$} 
     \addplot[blue, densely dashdotdotted, thick, smooth] table {\QLdetunppp};
    \addlegendentry{QL $\delta\theta =0.001$} 
        \addplot[black, thick, smooth] table {\CLHetphNum};
    \addlegendentry{CL heterodyne} 
    \addplot[black, dashed, thin, smooth] table {\CLULphNum};
    \addlegendentry{CL ultimate limit} 
  \end{axis}
  \node at (-1, 3) {a)};
\end{tikzpicture}
\begin{tikzpicture} 
    \begin{axis}[
    scale only axis, 
      width=0.8\columnwidth, 
      height=0.35\columnwidth, 
    ymode=log,
    xmin = 0.1,
    xmax=1, 
    xlabel={path transmissivity $\kappa$},
    legend style={font=\tiny, row sep=-0cm},
    legend pos=south west, 
     legend cell align={left}
    ]
  ]
    \addplot[blue, thick, smooth] table {\OptSqDatap};
    \addlegendentry{QL}
    \addplot[black, thick, smooth] table {\CLhetloss};
    \addlegendentry{CL heterodyne}
    \addplot[black, dashed , smooth] table {\CLULloss};
    \addlegendentry{CL ultimate limit}
  \end{axis}
  \node at (-1, 3) {b)};
\end{tikzpicture}
\caption{\label{fig:plot} Plotted is the performance metric $\text{Var}[\tau]\text{Var}[\omega]\simeq F^{-1}_{\hphantom{-1} \tau \tau}F^{-1}_{\hphantom{-1} \omega\omega}$ of the joint estimation task for the QL and CL protocols.  In (a) we fix transmissivity $\kappa=1$ and consider the QL with and without phase detunings. In (b) we fix photon number $N=100$, and optimize allocation of squeezing and displacement  on each of the three squeezed modes for each $\kappa$ with a  maximum squeezing level of $20$dB ($r\leq 2.31$)  at $\kappa=1$.}
\end{figure*}
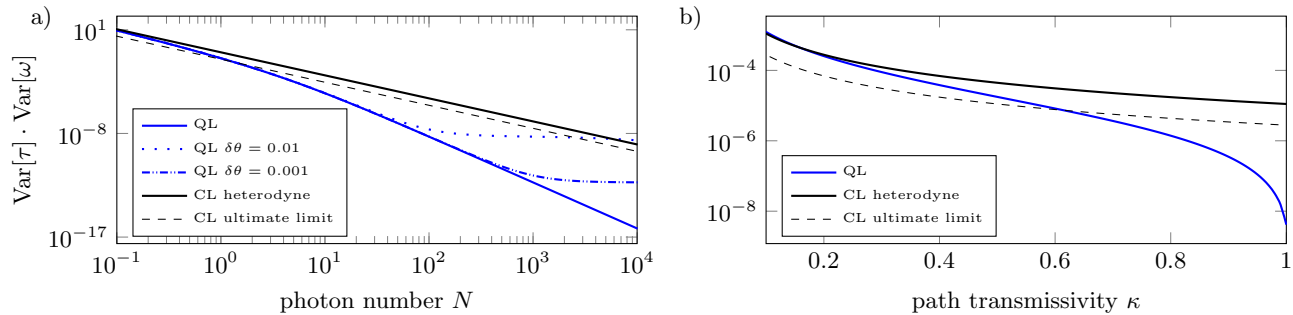

{\em Comparison and photon loss resilience.---}
Let us now compare the optimal performance of the QL and the CL. The relevant resource parameters that appear in the CRBs are the signal photon number $N$, the time duration $\Delta T$ of the pulse  and its bandwidth $\Delta\omega$. To guarantee a fair comparison, we set each of these parameters equal for the QL and CL. The time delay $\tau$ that also appears in the FIMs will be  set $\tau = 0$ for simplicity. We will quantify the performance of the joint estimation task with the squared geometrical mean of MSEs, i.e. $\text{Var}[\tau]\text{Var}[\omega]\simeq F^{-1}_{\hphantom{-1}\tau\tau}F^{-1}_{\hphantom{-1}\omega\omega}$. In the ideal case with no detunings and no photon loss, the QL achieves  $\text{Var}[\tau]\text{Var}[\omega]\sim N^{-4}$ and clearly outperforms the CL that only attains $\text{Var}[\tau]\text{Var}[\omega]\sim N^{-2}$. In Fig.~\ref{fig:plot} a), we plot $\text{Var}[\tau]\text{Var}[\omega]$  against photon number $N$ for the QL (with $N_{\text{sq}} = 3N/4$) and the CL. We observe that quantum advantage is achieved for $N>1.1$ ($r>0.5$) against the ultimate (potentially unattainable) classical limit.  Note that $r\approx 1$ is currently experimentally feasible \cite{presutti}.  Besides the ideal scenario, we also plot the QL performance with detunings and see that the Heisenberg scaling is lost around $N\approx 1/\delta\theta$. 
Let us now compare the performances  when photon loss is present, by  plotting $\text{Var}[\tau]\text{Var}[\omega]$ against the path transmissivity $\kappa$ in Fig.~\ref{fig:plot} b). We assumed no detunings for the QL and we optimized the allocation of displacement  and squeezing for each $\kappa$. We see that our optimized QL performs always better than the best-known CL, and achieves significant quantum advantage for $\kappa > 0.5$ even against the classical ultimate limit. 
In SM~\ref{app:numerical} we also considered heterodyne reception ($\delta \omega\Delta T\gg 1 $ and arbitrary $\delta\theta$) for the QL, and found a performance close to the best-known CL protocol. We found for some squeezing configurations a constant factor quantum advantage for single parameter estimation against the best-known CL for high round-trip transmissivities.

{\em Practical considerations of energy constraint.---} In the above comparison, we have set the total energy to be a resource constraint for a fair theoretical comparison. However, in practice, laser lidars can easily adopt a power on the order of 10 watts, which corresponds to $N_{\text{coh}}=10^7$ per mode at $1550$nm and $10$THz bandwidth; while squeezed vacuum sources generally have photon number $<10^3$. Indeed, a practical lidar scenario  is not constrained by total power---squeezing power and laser power (displacement) are constrained on their own due to device, which is the case in recent experiments~\cite{xia2020} and adopted in theory works~\cite{shi2023dualcomb}. Therefore, we consider in SM~\ref{app:displacementdominant} a case where $N_{\text{coh}} \gg N_{\text{sq}}$. In this case and when $N_{\text{sq}}\gg \kappa/(1-\kappa)$, we identify a constant advantage from squeezing, and the advantage is affected by loss in similar manner as the squeezing variance  $(\kappa e^{-2r}+(1-\kappa))/2$  is increased by a decreasing transmissivity $\kappa<1$.  We observe a similar loss behaviour in Fig.~\ref{fig:plot} b). In the ideal case $\kappa = 1$, we obtain an enhancing factor of $e^{-2r}$ for each parameter. Current technology can achieve $r\simeq 1$ \cite{presutti}, and thus one could optimistically expect an advantage of $F^{-1}_{\text{C,het} \, \tau \tau}  F^{-1}_{\text{C,het} \, \omega \omega}/F^{-1}_{\hphantom{-1}\tau\tau}F^{-1}_{\hphantom{-1}\omega\omega} \simeq 50$. 

{\em Conclusion.---}
In conclusion, we have proven that the Heisenberg limit is simultaneously attainable for the estimation of the range and velocity of a moving target with an experimentally feasible pulse of displaced squeezed states and homodyne detection. In the presence of loss, the quantum lidar always performs better than the best-known classical lidar, and obtains significant quantum advantage for round-trip transmissivities above $50\%$. The quantum advantage persists even in practical scenarios where quantum engineering is applied on top of a strong coherent state with power in watts.
This represents a notable step forward in the field, given that earlier protocols relied on experimentally complex states and/or measurement, and forfeit their quantum advantage with the loss of just one photon.

\begin{acknowledgments}
We thank G\"oran Johansson, Roberto Di Candia and Robert S. Jonsson for insightful discussion during the beginning of this project. The authors acknowledge support from EU FET Open project EPIQUS (899368) and HORIZON-CL4- 2022-QUANTUM01-SGA project 101113946 OpenSuperQPlus100 of the EU Flagship on Quantum Technologies, the Spanish Ram\'on y Cajal Grant RYC-2020-030503-I, project Grant No. PID2021-125823NA-I00 funded by MCIN/AEI/10.13039/501100011033 and by “ERDF A way of making Europe” and “ERDF Invest in your Future”, and from the IKUR Strategy under the collaboration agreement between Ikerbasque Foundation and BCAM on behalf of the Department of Education of the Basque Government. We acknowledge funding from Basque Government through Grant No. IT1470-22 and the IKUR Strategy under the collaboration agreement between Ikerbasque Foundation and BCAM on behalf of the Department of Education of the Basque Government.
M.R. acknowledges support from UPV/EHU PhD Grant PIF21/289. 
QZ acknowledges NSF CAREER Award CCF-2240641, National Science Foundation OMA-2326746, ONR Grant No. N00014-23-1-2296, Cisco Systems, Inc. and Halliburton Company.
\end{acknowledgments}

\appendix
\newpage 
\begin{widetext}

\section{Temporal mode operators}\label{app:Modes}
We use Hermite-Gaussian (HG) modes defined by
\begin{align}
    \phi_n^{\text{in}} (t)=\Phi_n  (t;\tau_0,\omega_0,\theta_0) = \left(\frac{\sigma^2}{2} \right)^{1/4} \sqrt{\frac{2^{-n}}{n! \sqrt{\pi}}}   H_n\left(\frac{\sigma
   (t-\tau_0 )}{\sqrt{2}}\right) e^{-\frac{1}{4} \sigma
   ^2 (t-\tau_0 )^2-i (\omega_0  t +\theta_0)} ,
\end{align}
where $H_n (x)$ are the physicist's Hermite polynomials. The modes above are the in modes. The out modes are obtained by changing the parameters $\tau_0\rightarrow \tau$, $\omega_0\rightarrow \omega$ and $\theta_0\rightarrow \theta$. Often, we will  omit the upper in/out label, as it will be clear from context to which of the modes we refer to. We also define the mode envelope $\bar{\phi}_n (t)$ as $\phi_n (t) = \bar{\phi}_n (t)e^{-i(\omega t +\theta)}$. We use the properties $ 2xH_{n} (x)  =  H_{n+1}(x)+2nH_{n-1}(x)  $ and $\partial_x H_{n} (x) = 2n H_{n-1} (x)$
to obtain after some manipulations for the out modes (omitting the out label)
\begin{align}
    \label{eq:partialtau} \partial_\tau \phi_n (t) &= -\frac{\sigma}{\sqrt{2}} \left( \sqrt{\frac{n}{2}} \phi_{n-1} (t) -\sqrt{\frac{n+1}{2}} \phi_{n+1}(t) \right) \\ \label{eq:partialomega}
     \partial_\omega \phi_n (t) &= - i\frac{\sqrt{2}}{\sigma} \left( \sqrt{\frac{n}{2}} \phi_{n-1} (t) +\sqrt{\frac{n+1}{2}} \phi_{n+1}(t) \right) - i \tau \phi_n (t)\\  \label{eq:partialtheta}
     \partial_\theta \phi_n (t) &= -i \phi_n (t) .
\end{align}
We note, that using the properties of the Hermite polynomials above we find $\phi_n (t) \cdot (t-\tau) = \frac{1}{\sigma} (\sqrt{n} \phi_{n-1}(t)+ \sqrt{n+1} \phi_{n+1}(t))$ which can be used to derive  Eq.~\eqref{eq:partialomega}.
As we can see, the derivatives with respect to $\tau$ and $\omega$ of mode $n$ have a similar structure in that they are a linear combination of neighbouring modes $n-1$ and $n+1$. We exploit this structure to attain the Heisenberg limit for both parameters simultaneously.
The (out) mode operators $\hat{a}_n^\dag = \int\mathrm{d}t\, \phi_n (t) \hat{E}^\dag(t)$ directly inherit the properties of the modes
\begin{align}
        \partial_\tau \hat{a}_{n}^\dag (t) &= -\frac{\sigma}{\sqrt{2}} \left( \sqrt{\frac{n}{2}} \hat{a}_{n-1}^\dag -\sqrt{\frac{n+1}{2}} \hat{a}_{n+1}^\dag \right) \\
     \partial_\omega \hat{a}_{n}^\dag &= - i\frac{\sqrt{2}}{\sigma} \left( \sqrt{\frac{n}{2}} \hat{a}_{n-1}^\dag +\sqrt{\frac{n+1}{2}} \hat{a}_{n+1}^\dag \right) - i \tau \hat{a}_{n}^\dag\\
     \partial_\theta \hat{a}_n^\dag &= -i \hat{a}_n^\dag .
\end{align}
The expansion coefficients $\gamma_{\alpha kn}$, defined via $\partial_\alpha \hat{a}_k^\dag = \sum_{n=0}^\infty \gamma_{\alpha kn} \hat{a}_n^\dag$, are for our parameters given by
\begin{align}
    \gamma_{\tau k n} &= -\frac{\sigma}{2} \left( \sqrt{k} \delta_{n ,k-1} - \sqrt{k+1} \delta_{n,k+1} \right) \\
    \gamma_{\omega k n} &=  -\frac{i}{\sigma} \left( \sqrt{k} \delta_{n ,k-1} + \sqrt{k+1} \delta_{n,k+1} \right) -i\tau \delta_{k,n} \\
    \gamma_{\theta k n} &= -i \delta_{k,n} .
\end{align}

\section{Displaced squeezed state lidar}\label{app:sq}
The class of probe states of the EM field we consider are Gaussian displaced squeezed states of multiple temporal modes~\cite{barnett}, whose wave function given in the time domain is
\begin{align}\label{eq:instate}
 \vert \psi^{\text{in}} \rangle = \hat{D} [s] \hat{S} [f] \vert 0 \rangle ,
\end{align}
where $\vert 0\rangle$ is the vacuum state and $\hat{D}[s] = \exp (\int \mathrm{d}t\, s(t) \hat{E}^\dag(t) -h.c. )$ and $\hat{S}[f] = \exp ( - \frac{1}{2}\int \mathrm{d} t \, \int \mathrm{d} t' \, f(t, t') E^\dag (t) E^\dag (t') + h.c. ) $ are the continuous-time versions of the standard displacement and squeezing operators with integration limits $-\infty$ to $+\infty$. 
The function $s(t)$ may be interpreted as a  complex-valued signal, analogously to the classical lidar literature~\cite{helstrom}. The complex-valued amplitude $f(t,t')$ describes  the quantum correlations of our state. Motivated by experimental considerations of SPDC pumped with a laser pulse \cite{wasilewski,roman2023,presutti} and algebraic convenience, we assume the form $f(t,t') = \sum_{n=0}^\infty \zeta_n \phi_n^\text{in} (t) \phi_n^\text{in}(t')$, where $\phi_n (t)$ are the Hermite-Gaussian modes. We additionally have $s(t) = \sum_n \alpha_n \phi_n^\text{in} (t)$.

Let us  review some properties of the displaced squeezed (in/out) state in the Hermite-Gaussian basis $\hat{D}\hat{S}\vert 0\rangle = \otimes_n \hat{D}_n (\alpha_n) \hat{S}_n (\zeta_n)\vert 0\rangle$ of  the main text that will be useful for the upcoming calculations.  For now we consider arbitrary mode configurations with arbitrary $\{\alpha_n \}_n$ and $\{ \zeta_n \}_n$. The squeezing and displacement operators have the transformation rules  given by
\begin{align}
    \hat{S}^\dag \hat{a}_n^\dag \hat{S} & = \hat{a}_n^\dag \cosh (  r_n ) - \hat{a}_n e^{-i\varphi_n} \sinh (r_n) \\
    \hat{S}^\dag \hat{a}_n \hat{S} &= \hat{a}_n \cosh (r_n) - \hat{a}_n^\dag e^{i\varphi_n} \sinh (r_n) \\
    \hat{D}^\dag \hat{a}_n \hat{D} &= \hat{a}_n + \alpha_n\\
     \hat{D}^\dag \hat{a}_n^\dag \hat{D} &= \hat{a}_n^\dag + \alpha_n^*  ,
\end{align}
where $\zeta_n = r_n e^{i \varphi_n}$ with $r_n \geq 0$.
With this, we can calculate the photon number of our general state $\hat{D}\hat{S}\vert 0\rangle = \otimes_n \hat{D}_n \hat{S}_n\vert 0\rangle$. The total photon number of the quantum state, using the mode decomposition $\hat{E}(t)=\sum_n \phi_n (t)\hat{a}_n$ and the above transformation rules, is given by
\begin{align}
  N=     \int_{-\infty}^\infty \langle  \hat{E}^\dag (t)\hat{E}(t)  \rangle &=  \sum_{n,m} \int_{-\infty}^\infty \phi_n^* (t) \phi_m (t) \langle 0 \vert \hat{S}^\dag \hat{D}^\dag \hat{a}_n^\dag \hat{a}_m \hat{D}\hat{S}\vert 0 \rangle 
  \\&= \sum_n \langle 0 \vert \hat{S}^\dag \hat{D}^\dag \hat{a}_n^\dag \hat{a}_n \hat{D}\hat{S}\vert 0 \rangle = \sum_{n=0}^\infty \sinh^2 (r_n) + \sum_{n=0}^\infty \vert \alpha_n \vert^2 .
 \end{align}
 As we can see, each mode $n$ contributes $\sinh^2 (r_n)$ photons due to squeezing and $\vert \alpha_n\vert^2$ due to displacement.
 For the photon number it does not matter whether we use the in or out modes and states as the process is photon number preserving in the ideal case.

Let us now come to the time duration $\Delta T$ and bandwidth $\Delta\omega$ of the signal. For a classical complex-valued signal $g(t)$, with total energy $N= \int \mathrm{d}t \vert g(t)\vert^2$, the time duration is naturally defined as~\cite{helstrom} $\Delta T^2 N = \int \mathrm{d}t \vert g(t)\vert^2 t^2 - (\int \mathrm{d}t \vert g(t)\vert^2 t)^2$ and the bandwidth is $\Delta \omega^2 N = \int \mathrm{d}\nu \vert \tilde{g}(\nu)\vert^2 \nu^2 - (\int \mathrm{d}t \vert \tilde{g}(\nu)\vert^2 \nu)^2$, where $\tilde{g}(\nu)$ is the Fourier transform of $g(t)$. Thus, these quantities are the variances of the energy distribution in time and frequency. To obtain the quantum analogue, we simply replace $\vert g(t)\vert^2$ with $\langle \hat{E}^\dag (t) \hat{E}(t)\rangle$ and $\vert \tilde{g}(\nu)\vert^2$ with $\langle \hat{A}^\dag (\nu) \hat{A}(\nu)\rangle$. So let us calculate these quantities. We start with the time duration. First, we calculate the normalized received energy as a function of time, using abbreviations $s_n=\sinh(r_n)$ and $c_n = \cosh (r_n)$
\begin{align}
    \frac{\langle \hat{E}^\dag (t) \hat{E}(t) \rangle }{N} &= \frac{\sum_{n,m} \phi_n^*(t) \phi_m (t) \langle 0 \vert \hat{S}^\dag \hat{D}^\dag \hat{a}_n^\dag \hat{a}_m \hat{D}\hat{S}\vert 0 \rangle}{N} \\
    &= \frac{\sum_{n} \vert \phi_n (t)\vert^2 s_n^2 +\sum_{n,m} \phi_n^* (t) \phi_m (t) \alpha_n^* \alpha_m}{N} \\
    &= \frac{\sum_{n} \vert \phi_n (t)\vert^2 s_n^2 +   \vert s(t) \vert^2}{N} ,
\end{align}
where we again used the transformation rules and $s(t)= \sum_n \alpha_n \phi_n (t)$. We already see that by setting the squeezing to $0$, i.e. $s_n=0$ for all $n$, we recover the result $\vert s(t)\vert^2/N$ from classical literature.
Let us now calculate the center time of our signal
\begin{align}
  \int_{-\infty}^\infty \mathrm{d}t \,  t\cdot \frac{\langle \hat{E}^\dag (t) \hat{E}(t) \rangle }{N}   =  \frac{\int\mathrm{d}t\, \sum_{n} t \cdot \vert \phi_n (t)\vert^2 s_n^2 + \int \mathrm{d}t  \, t \vert s(t) \vert^2}{N} =\tau,
\end{align}
which follows from the properties of our HG modes and from the fact that we assumed $s(t)$ to be centered around $\tau$, i.e. $\int\mathrm{d}t\, t \vert s(t)\vert^2 =\tau$. This will be true, if we prepare the coherent (in) signal at time $\tau_0$ at the beginning of the experiment. With this, we can directly calculate the time duration $\Delta T$ via  
\begin{align}
    \Delta T^2 &= \int_{-\infty}^\infty  \mathrm{d}t \,  (t-\tau)^2\cdot \frac{\langle \hat{E}^\dag (t) \hat{E}(t) \rangle }{N}   =  \frac{\int\mathrm{d}t\, \sum_{n} (t-\tau)^2\cdot \vert \phi_n (t)\vert^2 s_n^2 + \int \mathrm{d}t  \, (t-\tau)^2 \vert s(t) \vert^2}{N} \\
     &=   \frac{\int\mathrm{d}t\,  \sum_n \frac{2}{\sigma^2}\left( \sqrt{\frac{n}{2}} \bar{\phi}_{n-1} (t) + \sqrt{\frac{n+1}{2}} \bar{\phi}_{n+1} (t) \right)^2   s_n^2 + \int \mathrm{d}t  \, (t-\tau)^2 \vert s(t) \vert^2}{N}  \\
     &= \frac{ \sum_n \frac{2}{\sigma^2}\left( n+\frac{1}{2} \right)  s_n^2 }{N} +\frac{ \int \mathrm{d}t  \, (t-\tau)^2 \vert s(t) \vert^2}{N} ,
\end{align}
where we again used properties of the HG modes. We see that the time duration is composed of two parts, one coming from squeezing, the other from displacement.

Let us now come to the bandwidth. We start with
\begin{align}
  \frac{\langle \hat{A}^\dag(\nu) \hat{A}(\nu) \rangle }{N} & = \frac{1}{N} \frac{\int \mathrm{d}t}{\sqrt{2\pi}}\frac{\int \mathrm{d}t'}{\sqrt{2\pi}}  e^{-i\nu t} e^{+i\nu t'} \langle \hat{E}^\dag(t) \hat{E}(t') \rangle \\
    &=  \frac{\sum_{n} \tilde{\phi}_n^* (\nu) \tilde{\phi}_n (\nu)  s_n^2 +\sum_{n,m} \tilde{\phi}_n^* (\nu) \tilde{\phi}_m (\nu) \alpha_n^* \alpha_m}{N}  \\
    &= \frac{\sum_{n} \vert \tilde{\phi}_n (\nu) \vert^2  s_n^2 +\vert \tilde{s}(\nu)\vert^2}{N}  ,
\end{align}
where $\tilde{\phi}_n (\nu) = \int \mathrm{d}t e^{i\nu t} \phi_n (t)/\sqrt{2\pi}$ is the Fourier transform of our temporal mode function and $\tilde{s}(\nu)$ is the Fourier transform of $s(t)$. We have
\begin{align}
    \tilde{\phi}_n (\nu) \sim \left(\frac{2}{\sigma^2} \right)^{1/4} \sqrt{\frac{2^{-n}}{n! \sqrt{\pi}}}   H_n\left(\frac{
   \sqrt{2}(\nu-\omega )}{\sigma}\right) e^{-\frac{1}{\sigma^2}  (\nu-\omega )^2 } ,
\end{align}
where the the only difference with our temporal modes is the change $t\rightarrow \nu$, $\tau\rightarrow\omega$ and $\sigma^2/2\rightarrow 2/\sigma^2$. There are also some phase terms which we have omitted, as they are not important for the calculations. Due to this analogy, the calculations for the bandwidth are identical to the time duration calculations above and we find  
\begin{align}
    \Delta\omega^2 =\frac{ \sum_n \frac{\sigma^2}{2}\left( n+\frac{1}{2} \right)  s_n^2 }{N} +\frac{ \int \mathrm{d}\nu  \, (\nu-\omega)^2 \vert \tilde{s}(\nu) \vert^2}{N}  ,
\end{align}
assuming  the reflected signal $\tilde{s}(\nu)$ to be centered around $\omega$, which will be the case if the initial signal is centered around $\omega_0$.
For the state considered in the main text with the first three modes squeezed equally, i.e. $r_0 = r_1 = r_2 \equiv r$, and the $n=1$ mode displaced with arbitrary $\alpha_1$, while the rest of the modes are not populated, we obtain
\begin{align}
    \Delta \omega^2 = \frac{\sigma^2}{2} \frac{3}{2}   \quad\quad\quad\quad \Delta T^2 =  \frac{2}{\sigma^2} \frac{3}{2} .
\end{align}
Notably, the time duration and bandwidth do not depend on the relative allocation of squeezing and displacement for this particular mode configuration. 
The time-bandwidth product is
\begin{align}
    \Delta T  \Delta \omega  = \frac{3}{2} .
\end{align}
We can increase the time-bandwidth product by populating modes whose index is shifted as $n\rightarrow n+K$, i.e. $r_K = r_{K+1} = r_{K+2} \equiv r$, and the mode $K+1$ displaced with arbitrary $\alpha_{K+1}$. This yields $\Delta T \Delta\omega = K+3/2$.

\section{Homodyne measurement}\label{app:homodyne}
The homodyne detection works as follows. The received electric field, whose positive-part operator is $\hat{E}(t)$, is mixed in a balanced beam splitter with a local oscillator field  $\hat{E}_{\text{LO}}(t)$ giving rise to the new operator (in the Heisenberg picture)
\begin{align}
    \hat{E}_\pm (t) = \frac{\hat{E}(t) \pm \hat{E}_{\text{LO}}(t)}{\sqrt{2}}  .
\end{align}
We can measure the corresponding photon currents $ \hat{E}_\pm^\dag (t)  \hat{E}_\pm (t) $ and subtract them from each other yielding 
\begin{align}
    \hat{E}_+^\dag (t)  \hat{E}_+ (t) -\hat{E}_-^\dag (t)  \hat{E}_- (t) = \hat{E}(t) \hat{E}_{\text{LO}}^\dag(t) + \hat{E}^\dag(t) \hat{E}_{\text{LO}}(t) .
\end{align}
Assuming the state belonging to the operator $\hat{E}_{\text{LO}}(t)$ is in a coherent state with temporal mode function $ \sim e^{-i ( \omega_{\text{LO}} t+\theta_{\text{LO}})}$ satisfying the strong local-oscillator condition~\cite{shapiro2009}, we can replace the operators $\hat{E}_{\text{LO}}(t)$ by their expectation value and obtain the quadrature measurement operator
\begin{align}
    \hat{X}(t) = \frac{\hat{E}(t) e^{i \omega_{\text{LO}}t+i\theta_{\text{LO}}} +\hat{E}^\dag(t) e^{-i \omega_{\text{LO}}t -i\theta_{\text{LO}}} }{\sqrt{2}} .
\end{align}
The measurement outcomes $X(t)$ of this operator are time resolved and proportional to the measured  homodyne current. We now want to find the eigenstates of this operator. For this, we will expand the field operators in modes $\hat{E} (t) = \sum_n \phi_n^{\text{LO}} (t) \hat{a}_n^{\text{LO}}$, where we define $\phi_n^{\text{LO}} (t) = \Phi_n (t;\tau,\omega_{\text{LO}},\theta_{\text{LO}})$ and $\hat{a}^{\text{LO}}_n = \int \mathrm{d}t \, \phi_n^{\text{LO}*}(t) \hat{E}(t)$. We call this basis the LO basis, as its center frequency and phase equal the ones of the LO. We obtain
\begin{align} \label{eq:quadraturedecomp}
    \hat{X} (t) = \sum_{n=0}^\infty \bar{\phi}_n^\text{LO} (t)  \frac{  \hat{a}_n^\text{LO} +  \hat{a}_n^{\text{LO}\dag}}{\sqrt{2}},
\end{align}
where $\bar{\phi}_n^{\text{LO}}$ is the envelope of the LO mode defined via $\phi_n^{\text{LO}} (t) = \bar{\phi}_n^{\text{LO}} (t)e^{-i\omega_{\text{LO}}t-i\theta_{\text{LO}}}$.
Let us define $\hat{X}_{\text{LO},n} := (  \hat{a}_n^\text{LO} + \hat{a}_n^{\text{LO}\dag})/\sqrt{2}$. Apparently, measuring the quadratures $\hat{X}_{\text{LO},n}$ for all $n$ is equivalent to measuring the operator $\hat{X}(t)$ for all times. The eigenstates of the operators $\hat{X}_{\text{LO},n} $ are well known and given by~\cite{barnett}
\begin{align}
    \vert x_{\text{LO}, n}\rangle_n =   \pi^{-1/4} \exp \left( -\frac{1}{2} x_{n}^2 + \sqrt{2}   x_{n} \hat{a}_n^{\text{LO}\dag} -\frac{1}{2}  ( \hat{a}_n^{\text{LO} \dag } )^2 \right) \vert0\rangle_n ,
\end{align}
with eigenvalue $x_n$ and where we defined the vacuum $\vert 0\rangle_n$ of mode $n$.
The general eigenstate is then given by
\begin{align} \label{eq:eigenstatemode}
    \vert x_{\text{LO}}\rangle := \bigotimes_n \vert x_{\text{LO}, n}\rangle_n  = \left(\prod_n \pi^{-1/4} \right) \exp \left( -\frac{1}{2} \sum_n x_{n}^2 + \sqrt{2}   \sum_n  x_{n} \hat{a}_n^{\text{LO}\dag} -\frac{1}{2}  \sum_n (\hat{a}_n^{\text{LO}\dag})^2  \right) \vert0\rangle .
\end{align}
 In Ref.~\cite{serafini}, it was shown that if $\vert \psi\rangle$ is a Gaussian state, then $\vert \langle x_{\text{LO}} \vert \psi\rangle \vert^2$ is a Gaussian probability distribution of measurement outcomes $x_n$ with mean $\langle \hat{X}_{\text{LO},n}\rangle$ and covariance matrix $\langle \hat{X}_{\text{LO},n} \hat{X}_{\text{LO},m}\rangle$. As can be seen from Eq.~\eqref{eq:quadraturedecomp}, we have $x_n = \int \mathrm{d}t\, X(t) \bar{\phi}_n^\text{LO} (t)$ and conversely $X(t) = \sum_n x_n \bar{\phi}_n^\text{LO} (t)$. In principle, our index $n$ goes to infinity, but we assume that for practical purposes, we can consider $n$ to assume finitely many values.  We similarly assume in the time basis, that it is suffices to consider finite measurement times.

Let us now reformulate the above result in the time basis.  For this, let us consider the terms inside the exponential in Eq.~\eqref{eq:eigenstatemode}. First
\begin{align}
    \sum_n x_n \hat{a}_n^{\text{LO}\dag} = \int \mathrm{d}t \, \sum_n x_n \bar{\phi}_n^\text{LO} (t)  e^{-i\omega_{\text{LO}}t-i\theta_\text{LO} } \hat{E}(t) =  \int \mathrm{d}t \, x(t) e^{-i\omega_{\text{LO}}t-i\theta_\text{LO} } \hat{E}(t) .
 \end{align}
 Now the second relevant term is
\begin{align}
    \sum_n (\hat{a}_n^{\text{LO}\dag })^2= \int \mathrm{d}t \int \mathrm{d}t' \sum_n \bar{\phi}_n^\text{LO}(t) \bar{\phi}_n^\text{LO}(t') e^{-i\omega_{\text{LO}}(t+t')-2i\theta_\text{LO}} \hat{E}(t)\hat{E}(t') = \int\mathrm{d}t\, e^{-i2\omega_{\text{LO}}t-2i\theta_\text{LO}} \hat{E}(t)^{\dag 2} ,
\end{align}
using completeness relation $\sum_n \bar{\phi}_n^\text{LO} (t) \bar{\phi}_n^\text{LO}(t') =\delta (t-t')$. We also note $\sum_n x_n^2 =\int\mathrm{d}t \, X^2(t)$. With this we find
\begin{align}
     \vert x_{\text{LO}}\rangle= \left(\prod_n \pi^{-1/4} \right) \exp \left(\int\mathrm{d}t \left[  -\frac{1}{2} X^2(t) + \sqrt{2}    X(t) e^{-i\omega_{\text{LO}}t -i  \theta_{\text{LO}} } \hat{E}(t) -\frac{1}{2}    e^{-i2\omega_{\text{LO}}t-2i\theta_{\text{LO}}} \hat{E}(t)^{\dag 2} \right] \right) \vert0\rangle  ,
\end{align}
which is the same result as above just in the time basis.

The next task is to obtain the statistics of time bin measurement operators $\hat{X}_i =  \int_{i\Delta t}^{(i+1)\Delta t} \mathrm{d}t \, \hat{X}(t)/\Delta t$ of arbitrary time length $\Delta t$. First, we define time bin operators $\hat{E}_{\delta t,i} = \int_{i\delta t}^{(i+1)\delta t} \mathrm{d}t\hat{E}(t)/\delta t$ of infinitesimal short length $\delta t \rightarrow 0$. We assume it to be a valid approximation to replace $\hat{E}(t)$ with $\hat{E}_{\delta t, i}$, and replace the integral $\int \mathrm{d}t \rightarrow \sum_i \delta t$ in all the expressions, as long as $\delta t$ is chosen small enough. We assume the index to be an element of a finite set of consecutive integers, implying the total measurement time to be finite. We assume this to give accurate results as long as the signal is fully contained within our measurement time. Let us consider the time bin version $\hat{X}_{\delta t, i}$ of the homodyne operator. The measurement outcomes $X_{\delta t ,i}\simeq X(i \delta t )$ are then Gaussian distributed with mean $\langle \hat{X}_{\delta t,i}\rangle$ and covariance matrix $\langle \hat{X}_{\delta t,i} \hat{X}_{\delta t,j}\rangle$, using the same arguments as above. Let us now consider larger disjoint time bins of length $(L+1)\delta t = \Delta t$ that are composed of our limitingly small time bins, that is $\hat{X}_k = \int_{k\Delta t}^{(k+1)\Delta t} \mathrm{d}t \, \hat{X}(t)/\Delta t \simeq (\hat{X}_{\delta t, k\cdot (L+1)} + \hat{X}_{\delta t, k\cdot (L+1)+1} +\ldots +\hat{X}_{\delta t, k\cdot (L+1)+L})/(L+1)\delta t$, where $k$ is an element of a finite set of consecutive integers (again implying a finite measurement time).  We can show by exploiting Gaussianity of the limitingly short time bins, that the statistics of the disjoint longer time bin data $X_i$ remains Gaussian distributed with mean $\langle \hat{X}_k\rangle$ and covariance matrix $\langle\hat{X}_k\hat{X}_q\rangle$.

This is derived from the following. A random vector $\Vec{y}\in \mathds{R}^N$  is distributed according to  a multivariate Gaussian distribution, if and only if $\vec{q} \cdot\Vec{y}$   is distributed according to  a univariate Gaussian distribution for all choices of  constant vectors  $\vec{q}\in \mathds{R}^N$ Ref.~\cite{Rao}. Note, that the special case $\vec q = 0$ will lead to a distribution with zero variance which we also consider to be Gaussian.   Suppose $\Vec{y}\in \mathds{R}^N$ is distributed according to  a multivariate Gaussian distribution. This means $\vec{q} \cdot\Vec{y}  = q_1 y_1+ \ldots q_N y_N  $   is distributed according to a univariate Gaussian distribution.
 Let us now consider the pair of random variables $\vec x = (x_1 , x_2)^T=(\vec{u} \cdot\Vec{y},\vec{v} \cdot\Vec{y})^T$, where $\Vec{u},\vec{v} \in \mathds{R}^N$ are constant vectors. This pair of random variables is distributed according to a bivariate Gaussian distribution which we will show in the following. For each choice of constant vector $\vec w \in\mathds{R}^N$ we have that 
\begin{align}
    \vec w \cdot \vec x = w_1 x_1 + w_2 x_2 &= w_1 \left( u_1 y_1 + \ldots + u_N y_N \right) + w_2 \left( v_1 y_1 + \ldots + v_N y_N \right) \\
    &= \left( w_1 u_1 + w_2 v_1 \right) y_1 + \ldots +  \left( w_1 u_N + w_2 v_N \right) y_N
\end{align}
is distributed according to a univariate Gaussian distribution, which directly follows from our initial assumption.
The above can now be extended to arbitrary linear combinations $(\vec{u_1}\cdot \vec{y},  {\vec{u_2}\cdot \vec{y}}, \vec{u_3}\cdot \vec{y},\ldots )$. In our case, each $\vec{u}_i$ represents a projection onto one of the disjoint long time bins. Thus, also the long time bin homodyne operators $\hat{X}_i$ are multivariate Gaussian distributed.

 \section{Quantum Fisher information matrix for mode parameters} \label{app:QFIMmode}
For general states of the form $\vert \psi^{\text{out}} \rangle = \sum_{N_0,N_1,\ldots} c_{N_0,N_1,\ldots} \frac{\hat{a}_0^{\text{out}\dag {N_0}}}{\sqrt{N_0!}} \, \frac{\hat{a}_1^{\text{out} \dag {N_1}}}{\sqrt{N_1!}} \ldots \vert 0 \rangle$, where only the mode operators $\hat{a}_n^{\text{out}\dag}$ depend on the parameters of interest, not the coefficients $c_{N_0,N_1,\ldots}$, the QFIM is given by Ref.~\cite{gessner2023}
\begin{equation}
   J_{\alpha\beta} = 4\text{Re} \left[\langle  \psi \vert \hat{H}_\alpha^\dag \hat{H}_\beta \vert \psi \rangle + \langle\psi \vert \hat{H}_\beta\vert \psi\rangle \langle\psi \vert \hat{H}_\alpha^\dag\vert \psi\rangle   \right]  ,
\end{equation}
where $\hat{H}_\alpha = \sum_{n=0}^\infty \left(\partial_{\alpha} \hat{a}_n^\dag\right) \hat{a}_n $ is an effective beam splitter Hamiltonian which couples each temporal out mode with its derivative, where we  omitted the out labels. It is shown, that the derivative of the state is given by~\cite{gessner2023}
\begin{align} \label{app:derivativeState}
    \partial_\alpha \vert \psi \rangle = \sum_{n=0}^\infty \left(\partial_{\alpha} \hat{a}_n^\dag\right) \hat{a}_n \vert\psi\rangle = \sum_{n=0}^\infty \sum_{k=0}^\infty \gamma_{\alpha n k} \hat{a}_k^\dag \hat{a}_n \vert \psi\rangle ,
\end{align}
with $\gamma_{\alpha nk}$ defined via $\partial_\alpha \hat{a}_n^\dag = \sum_k \gamma_{\alpha n k} \hat{a}_k^\dag$. 
As discussed in Ref.\cite{gessner2023}, to attain the Heisenberg limit, it is necessary to populate at least one mode and its derivative. In our particular case, this means $\hat{a}_n$ and $\hat{a}_{n+1}$ for example.

\section{Displaced squeezed state lidar QFIM}\label{app:sqQFIM}
We can now calculate the QFIM for  mode parameters $\mu_i =\{ \tau,\omega,\theta \}$. We  consider the state $\otimes_n [\hat{D}_n \hat{S}_n ]\vert 0\rangle$ with arbitrary $\{ \alpha_n \}_n$ and $\{ \zeta_n \}_n$.  As shown in Ref.~\cite{gessner2023}, we can write the derivative of the state as
\begin{align}
    \partial_{\mu_i} \vert \psi\rangle = \sum_{k=0}^\infty \left( \partial_{\mu_i} \hat{a}_k^\dag \right) \hat{a}_k \vert \psi\rangle =  \sum_{k=0}^\infty \left( \sum_{n=0}^\infty \gamma_{i k n} \hat{a}_n^\dag  \right) \hat{a}_k \vert \psi\rangle = \sum_{k,n} \gamma_{ikn} \hat{a}_n^\dag   \hat{a}_k \vert \psi\rangle ,
\end{align}
where the $\gamma$ coefficients were introduced in SM~\ref{app:Modes}. Here we used that the derivative mode can be expanded in terms of the regular modes. Let us again remind that we will use abbreviations $\bigotimes_{n=0}^\infty \hat{S}_n (\zeta_n) \equiv \hat{S}$, $\sinh (r_n) = s_n$, $\cosh (r_n) = c_n$ and $\bigotimes_{n=0}^\infty \hat{D}_n (\alpha_n) \equiv \hat{D}$. We find
\begin{align}
  \partial_{\mu_i} \vert \psi\rangle &=   \sum_{k,n} \gamma_{ikn} \hat{a}_n^\dag   \hat{a}_k \hat{D}\hat{S} \vert 0\rangle  = \hat{D} \sum_{k,n} \gamma_{ikn} \hat{D}^\dag \hat{a}_n^\dag \hat{D}\hat{D}^\dag \hat{a}_k \hat{D} \hat{S} \vert 0\rangle  \\
  &= \hat{D} \sum_{k,n} \gamma_{ikn} \left( \hat{a}_n^\dag + \alpha_{n}^*\right) \left( \hat{a}_k + \alpha_{k}\right) \hat{S}\vert 0\rangle \\
  &= \hat{D} \sum_{k,n} \gamma_{ikn} \hat{a}_n^\dag \hat{a}_k^\dag \hat{S}\vert 0\rangle + \hat{D} \sum_{k,n} \gamma_{ikn} \alpha_n^* \alpha_k \hat{S} \vert 0 \rangle + \hat{D} \sum_{k,n} \gamma_{ikn} (\alpha_n^* \hat{a}_k + \alpha_k \hat{a}_n^\dag) \hat{S}\vert 0 \rangle .
\end{align}
Let us now analyze each of the three terms appearing above. We start with the first term which is fully due to squeezing
\begin{align}
     \hat{D} \sum_{k,n} \gamma_{ikn} \hat{a}_n^\dag \hat{a}_k^\dag \hat{S}\vert 0\rangle  & = \hat{D} \hat{S} \sum_{k,n} \gamma_{ikn} \left( \hat{a}_n^\dag c_n - \hat{a}_n e^{-i\varphi_n} s_n \right)  \left( \hat{a}_k c_k - \hat{a}_k^\dag e^{+i\varphi_k} s_k \right) \vert 0 \rangle \\
    & = \hat{D}\hat{S} \sum_{k,n} \gamma_{ikn} \left( \hat{a}_n^\dag c_n - \hat{a}_n e^{-i\varphi_n} s_n \right)  \left(  -  e^{+i\varphi_k} s_k \right)\hat{a}_k^\dag \vert 0 \rangle \\
    &= -\hat{D} \hat{S} \sum_{k,n} \gamma_{ikn} e^{i\varphi_k} c_n s_k \hat{a}_n^\dag \hat{a}_k^\dag \vert 0 \rangle + \hat{D} \hat{S} \sum_{k} \gamma_{ikk} s_k^2\vert 0\rangle .
\end{align}
The second term $ \hat{D} \hat{S} \sum_{k,n} \gamma_{ikn} \alpha_n^* \alpha_k \vert 0 \rangle$ is fully due to displacement and does not need further simplifications.
The third term, which is mixed term between displacement and squeezing, is given by
\begin{align}
    \hat{D} \sum_{k,n} \gamma_{ikn} \alpha_n^* \alpha_k \hat{S} \vert 0 \rangle = \hat{D}\hat{S} \sum_{k,n} \gamma_{ikn} \left(  - \alpha_n^* s_k e^{i\varphi_k} \hat{a}_k^\dag + \alpha_k c_n \hat{a}_n^\dag\right) \vert 0\rangle .
\end{align}
We summarize
\begin{multline}
    \partial_{\mu_i}\vert \psi\rangle \\=  -\hat{D} \hat{S} \sum_{k,n} \gamma_{ikn} e^{i\varphi_k} c_n s_k \hat{a}_n^\dag \hat{a}_k^\dag \vert 0 \rangle + \hat{D} \hat{S} \Big( \sum_{k} \gamma_{ikk} s_k^2 +\sum_{k,n} \gamma_{ikn} \alpha_n^* \alpha_k \Big) \vert 0 \rangle + \hat{D}\hat{S} \sum_{k,n} \gamma_{ikn} \left(  - \alpha_n^* s_k e^{i\varphi_k} \hat{a}_k^\dag + \alpha_k c_n \hat{a}_n^\dag\right) \vert 0\rangle .
\end{multline}
Let us remind that the QFIM elements are given by
$
    J_{ij} = 4 \text{Re} \left[ \langle \partial_{\mu_i}\psi \vert \partial_{\mu_j}\psi \rangle  - \langle \partial_{\mu_i} \psi \vert \psi\rangle \langle \psi \vert \partial_{\mu_j} \psi\rangle\right]
$.
We have
\begin{align}
     \langle \partial_{\mu_i}\psi \vert \partial_{\mu_j}\psi \rangle &= \sum_{k'n'}\sum_{k,n} \gamma^*_{ik'n'}\gamma_{jkn} c_{n'} c_n s_{k'}s_k e^{i(\varphi_k-\varphi_{k'})} \left( \delta_{n' n} \delta_{k'k}+\delta_{nk'}\delta_{n'k}  \right) + \sum_{k'} \gamma^*_{ik'k'}s_{k'}^2 \sum_{k} \gamma _{jkk}s_{k}^2 \\
     &+ \sum_{k'n'}\sum_{k,n} \gamma_{i k' n'}^* \gamma_{jkn} \Big[  \alpha_{n'} \alpha_n^* s_{k'} s_k e^{i (\varphi_k -\varphi_{k'})} \delta_{k k'} - \alpha_{k'}^* \alpha_n^* c_{n'} s_k e^{i \varphi_k} \delta_{n' k} \\
     &\quad \quad \quad\quad\quad \quad \quad\quad \quad- \alpha_{n'} \alpha_k s_{k'} c_n e^{-i \varphi_{k'}} \delta_{k' n} + \alpha_{k'}^* \alpha_k c_{n'} c_n \delta_{n n'} \Big] .
\end{align}
With this, we finally arrive at 
\begin{align}
    J_{ij} = 4 \text{Re} &\Bigg[ \sum_{k,n} \gamma^*_{ikn} \gamma_{jkn} c_n^2 s_k^2 + \sum_{k,n} \gamma^*_{ink}\gamma_{jkn} c_k c_n s_k s_n e^{i(\varphi_k-\varphi_n)} \\
    &+\sum_{k,n,n'} \gamma_{ikn'}^* \gamma_{jkn} \alpha_{n'} \alpha_n^* s_k^2 - \sum_{k,n,k'} \gamma^*_{ik'k} \gamma_{jkn} \alpha_{k'}^* \alpha_n^* c_k s_k e^{i\varphi_k} \\
    &- \sum_{n,k,n'} \gamma^*_{inn'} \gamma_{jkn} \alpha_{n'} \alpha_k s_n c_n e^{-i \varphi_n} + \sum_{n,k,k'} \gamma^*_{ik'n} \gamma_{jkn} \alpha_{k'}^* \alpha_k c_n^2\Bigg] ,
\end{align}
noting that the vacuum terms cancelled out.
This is the general formula for the QFIM for the ideal case $\kappa =1$. We now calculate the QFIM with the previously given $\gamma_{ikn}$ for $\tau$, $\omega$ and $\theta$ and invert this matrix to obtain the corresponding QCRBs. For  the mode configuration considered in the main text $\hat{D}_1 (N_\text{coh}e^{-i\pi/4}) \hat{S}_0(r)\hat{S}_1 (re^{i\pi/2})\hat{S}_2(\hat{r})\vert 0 \rangle$, we obtain by inverting the matrix $J$ the result  $J^{-1}_{\hphantom{-1}\tau \tau} = \frac{1}{ \Delta \omega^2 N^2}$ and $J^{-1}_{\hphantom{-1}\omega\omega} =\frac{1}{  \Delta T^2   N^2 }$, where we optimized over the allocation of squeezing and displacement with the optimum $N_{\text{sq}}= 3N/4 $ assuming $N\gg 1$. This result is actually equal to the FIM of the same state and homodyne measurement as we will later show, implying optimality of homodyne detection at least for $\kappa = 1$.

\section{Coherent state lidar QFIM}\label{app:cohQFIM}
The classical benchmark lidar employs the coherent state $\hat{D}[s]\vert 0\rangle$. For now, we assume $s(t)$ to be of the form $s(t)= f(t-\tau)e^{-(i\omega t+\theta)}$ (after reflection), with a possibly complex-valued $f(t)$ with $\int\mathrm{d}t\, t\vert f(t)\vert^2 = 0$. The total number of photons is given by
\begin{align}
    N = \int_{-\infty}^\infty \mathrm{d}t\, \vert s(t)\vert^2 .
\end{align}
The definition of the time duration $\Delta T$ is given in Appendix~\ref{app:sq} and for a coherent state we have
\begin{align}
    \Delta T^2 \cdot N = \int_{-\infty}^\infty \mathrm{d}t\, t^2 \vert s(t)\vert^2 - \left( \int_{-\infty}^\infty \mathrm{d}t\, t \vert s(t)\vert^2 \right)^2 ,
\end{align}
which recovers the expression for classical signals in Ref~\cite{helstrom}. The expression for the bandwidth is given in Appendix~\ref{app:sq} and for coherent states we can show using properties of Fourier transforms  that 
\begin{align}
    \Delta \omega^2 \cdot N = \int_{-\infty}^\infty \mathrm{d}t\, \vert \partial_t s(t) \vert^2  - \left( i \int_{-\infty}^\infty \mathrm{d}t\, s^* (t) \partial_t s(t) \right)^2 ,
\end{align}
which recovers the expression for classical signals in Ref~\cite{helstrom}.

Let us now calculate the QFI for the coherent state. We use our (out) mode basis for the reflected signal $s(t) = \sum_k \alpha_k \phi_k (t)$, such that our state can be written as $\hat{D}\vert 0\rangle=\bigotimes_{n=0}^\infty \hat{D}_n\vert 0\rangle$, where $\hat{D}_n = \exp (\alpha_n \hat{a}_n^\dag-\alpha_n^* \hat{a}_n)$. We have the transformation rule $\hat{D}^\dag \hat{a}_n \hat{D} = \hat{a}_n + \alpha_n$ and $\hat{D}^\dag \hat{a}_n^\dag \hat{D} = \hat{a}_n^\dag + \alpha_n^*$. With this, we can calculate the QFI. We first calculate the derivative of the (out) state using Eq.~\eqref{app:derivativeState}  with respect to the parameters
\begin{align}
    \partial_{\mu_i} \vert \psi\rangle &= \sum_{k,n} \gamma_{ikn} \hat{a}_n^\dag \hat{a}_k \vert \psi\rangle = \hat{D}  \sum_{k,n} \gamma_{ikn}  \hat{D}^\dag\hat{a}_n^\dag \hat{D} \hat{D}^\dag \hat{a}_k \hat{D} \vert 0\rangle \\
    &= \hat{D} \sum_{k,n} \gamma_{ikn} \left( \hat{a}_n^\dag + \alpha_{n}^*\right) \left( \hat{a}_k + \alpha_{k}\right)\vert 0\rangle \\
    &= \hat{D} \sum_{k,n} \gamma_{ikn} \alpha_{k} \hat{a}_n^\dag \vert 0\rangle + \hat{D} \sum_{k,n} \gamma_{ikn} \alpha_{n}^* \alpha_k \vert 0\rangle .
\end{align}
We have
\begin{align}
    \langle \partial_{\mu_i} \psi\vert \partial_{\mu_j} \psi \rangle = \sum_{k',k,n} \gamma^*_{ikn}\gamma_{jkn} \alpha_{k'}^* \alpha_k + \sum_{k',n'} \gamma_{ik'n'}^* \alpha_{n'} \alpha_{k'}^* \sum_{k,n} \gamma_{jkn} \alpha_n^* \alpha_k
\end{align}
and
\begin{align}
    \langle\psi\vert \partial_{\mu_i} \psi\rangle = \sum_{k,n} \gamma_{ikn} \alpha_{n}^* \alpha_{k} .
\end{align}
We remind again that the QFIM is 
$
    J_{ij} = 4 \text{Re} \left[ \langle \partial_{\mu_i}\psi \vert \partial_{\mu_j}\psi \rangle  - \langle \partial_i \psi \vert \psi\rangle \langle \psi \vert \partial_j \psi\rangle\right] 
$. And we thus have
\begin{align}
      J_{ij} = 4 \text{Re} \left[  \sum_{k',k,n} \gamma^*_{ikn}\gamma_{jkn} \alpha_{k'}^* \alpha_k  \right] .
\end{align}
Let us relate this back to $s(t)$. We have 
\begin{align}
    \partial_{\mu_i} s(t) = \sum_k \alpha_k \partial_{\mu_i} \phi_k (t) =    \sum_k \alpha_k \sum_n \gamma_{ikn} \phi_n (t) = \sum_{k,n} \gamma_{ikn}  \alpha_k  \phi_n (t) .
\end{align}
For the overlap, we have
\begin{align}
    \int \mathrm{d} t \, \left(\partial_{\mu_i} s^* (t) \right) \left(\partial_{\mu_j} s (t) \right) &=  \int \mathrm{d} t \, \sum_{k',n'} \gamma_{ik'n'}^* \alpha^*_{k'} \phi_{n'}(t) \sum_{k,n} \gamma_{jkn} \alpha_k \phi_n (t) \\
    &= \sum_{k',k,n} \gamma_{ik'n}^* \gamma_{jkn} \alpha^*_{k'} \alpha_k .
\end{align}
So with this we have
\begin{align}
    J_{ij} = 4\text{Re} \left[  \int \mathrm{d} t \, \left(\partial_{\mu_i} s^* (t) \right) \left(\partial_{\mu_j} s (t) \right)  \right] .
\end{align}
We would like to relate these quantities to the time duration and bandwidth. We have
\begin{align}
    J_{\omega\omega} &= 4 \text{Re} \left[ \int\mathrm{d}t\, \partial_{\omega} s^*(t) \partial_{\omega} s(t) \right]  = 4 \text{Re} \left[ \int\mathrm{d}t\, \partial_{\omega} s^*(t) \partial_{\omega} s(t) \right] = 4 \text{Re} \left[ \int\mathrm{d}t\, i t s^*(t) (-i)t s(t) \right] \\
    &=  4 \text{Re} \left[ \int\mathrm{d}t\, t^2 \vert s(t)\vert^2 \right] = 4 \Delta T^2 N + 4 \tau^2 N
\end{align}
and
\begin{align}
     J_{\tau\tau} &= 4 \text{Re} \left[ \int\mathrm{d}t\, \partial_{\tau} s^*(t) \partial_{\tau} s(t) \right] = 4 \Delta \omega^2 N,
\end{align}
which follows from properties of the Fourier transform and the fact that we assumed $s(t) = f(t-\tau)e^{-i(\omega t +\theta)}$.
For the phase term we obtain $J_{\theta\theta} =  4 N$, $J_{\tau \theta} = 0$ and $J_{\omega \theta} = 4 \tau N$. Let us consider
\begin{align}
    J_{\omega \tau} = 4 \text{Re} \left[ \int\mathrm{d}t\, \partial_{\omega} s^*(t) \partial_{\tau} s(t) \right]  .
\end{align}
This term is $0$ when we assume our signal to be not frequency modulated, which implies the form $s(t) = \bar{s}(t-\tau) e^{-i (\omega t+\theta)}$, where $\bar{s}(t)$ is a real-valued function centered around $0$.  Thus, assuming no frequency modulation, yields the result given in the main text. This result was simply obtained by inverting the QFIM of the three parameters $\tau$, $\omega$ and $\theta$. Let us now briefly discuss the scenario, in which we have frequency modulation and $J_{\omega \tau} \neq 0$. By using Cauchy-Schwarz, we find $J_{\omega \tau}^2 \leq (4 \Delta\omega^2 N) \cdot (4\Delta T^2 N + 4 \tau^2 N)$. Assuming $\tau = 0$ for simplicity, we can show that the term $J^{-1}_{\hphantom{-1} \tau\tau} \cdot J^{-1}_{\hphantom{-1} \omega\omega}$ will always get larger (and thus worse), when $J_{\omega \tau} > 0$. Thus, frequency modulation is not helpful in our protocol at a fundamental level, and we will therefore not consider it. The same was found in classical literature~\cite{helstrom} for lidars with heterodyne detection.

\section{Coherent state lidar FIM for homodyne/heterodyne detection}\label{app:cohFIM}
Let us now calculate the FIM for the coherent state and the homodyne/heterodyne measurement $\hat{X} (t)$. We assume finite time resolution such that we measure time bin versions of the homodyne current $\hat{X}_i = \int_{i\Delta t}^{(i+1)\Delta t} \mathrm{d}t \hat{X}(t)/\Delta t$. We further assume for simplicity, that our (out) signal can be written as $s(t) = \bar{s}(t-\tau) e^{-i (\omega t + \theta)}$, where $\bar{s}(t)$ is the real-valued envelope. Assuming $\bar{s}(t-\tau)\cos (\delta\omega t + \delta\theta)$ does not vary much within each time bin of length $\Delta t$, we have for the mean $\mu_i \simeq  \sqrt{2} \Bar{s}(t_i-\tau) \cos (\delta\omega t_i +\delta\theta)$, where we defined $t_i = i\Delta t$, and the covariance matrix is $\Sigma_{ij} = \delta_{ij}/2\Delta t $. The FIM is now easily calculated using the formula for Gaussian distributions $ F_{\alpha \beta} = \partial_{\alpha} \vec{\mu} \cdot \Sigma^{-1} \cdot \partial_{\beta} \vec{\mu} +  \frac{1}{2} \text{Tr}\left[ \Sigma^{-1}\cdot \partial_\alpha \Sigma \cdot\Sigma^{-1} \cdot \partial_\beta \Sigma \right]$ \cite{VanTrees2001}, where only the first term contributes. We start with the time delay
\begin{align}
    F_{\tau\tau} &= 4 \sum_i \Delta t \, \left( \partial_\tau \bar s(t_i-\tau)\right)^2 \cos ^2(\delta\omega (t_i-\tau)+ \delta\omega \tau+\delta\theta) \\
    & \simeq 4 \int_{-\infty}^\infty \mathrm{d}t\, \left( \partial_\tau \bar s(t-\tau)\right)^2 \cos^2 (\delta\omega (t-\tau)+ \delta\omega \tau+\delta\theta) ,
\end{align}
where we switched back to continuous time which is valid due to our initial assumption that $\bar{s}(t-\tau)\cos (\delta\omega t + \delta\theta)$ does not vary much within each time bin. For the frequency, we get
\begin{align}
   F_{\omega\omega} &= 4 \sum_i \Delta t \, \bar{s}^2(t_i-\tau) t_i^2\sin  ^2(\delta\omega (t_i-\tau)+\delta\omega \tau+\delta\theta) \\
   & \simeq 4 \int_{-\infty}^\infty \mathrm{d}t\, \bar{s}^2(t-\tau) t^2\sin^2  (\delta\omega (t-\tau)+\delta\omega \tau+\delta\theta) .
\end{align}
Analogously, we can calculate the off-diagonal terms.
We call the measurement homodyne, if $\delta\omega \Delta T \ll 1$. In this case, the trigonometric functions can be approximated with small-angle approximations during the pulse duration $\Delta T$.  In this regime, the performance depends strongly on the phase detuning and we have to match the LO frequency and phase to the returned signal's carrier frequency and phase. This matching is possible, if sufficiently good prior information about the parameters is available. This prior knowledge may come from previous measurement runs or other sources. If $\delta\theta + \delta\omega \tau \approx 0$, we get $F_{\tau\tau} = 4 \Delta \omega^2 N$ and $F_{\omega\omega} = 0$. If $\delta\theta \simeq \pi/2-\delta\omega\tau$, we get $F_{\tau\tau} = 0$ and $F_{\omega\omega} = 4 (\Delta T^2 +\tau^2) N$.  So let us consider the intermediate scenario $\delta\theta = \pi/4 -\delta\omega\tau$. Here, we get $F_{\tau\tau} = 2\Delta \omega^2 N$, $F_{\omega\omega} = 2 (\Delta T^2 +\tau^2) N$, $F_{\theta\theta} = 2 N$, $F_{\tau\omega} = N$ (using partial integration), $F_{\tau\theta} = 0$ and $F_{\omega\theta}= 2 N \tau$. Inverting this matrix yields $F^{-1}_{\hphantom{-1}\tau\tau} =1/(2\Delta \omega^2 N (1-1/2\Delta T^2))$ and $F^{-1}_{\hphantom{-1}\omega \omega} =1/(2\Delta T^2 N (1-1/2\Delta \omega^2))$ with a product $F^{-1}_{\hphantom{-1}\tau\tau}\cdot F^{-1}_{\hphantom{-1}\omega \omega} = 4\Delta T^2 \Delta\omega^2/(4\Delta T^2\Delta\omega^2 N- N)^2$, which goes to infinity as $\Delta T^2\Delta\omega^2 \rightarrow 1/4$, where as we have  $F^{-1}_{\hphantom{-1}\tau\tau}\cdot F^{-1}_{\hphantom{-1}\omega\omega} = 1/ 4\Delta T^2\Delta\omega^2 N^2$ as $\Delta T^2 \Delta\omega ^2 \gg 1$. We thus see, that again, the performance is not optimal for the joint estimation task, i.e. it does not reach the QCRBs. Therefore, homodyne detection is not an optimal measurement for the joint estimation of range and velocity.

Now consider heterodyne reception, which we have when $\delta\omega \Delta T\gg 1$ and with arbitrary $\delta\theta$. In this case, the squared trigonometric functions oscillate rapidly under the integral. In this case, we can approximate the functions as $\sin^2 (\delta\omega t+\delta\theta) \simeq \cos^2 (\delta\omega t+\delta\theta) \simeq 1/2$ and $\sin  (\delta\omega t+\delta\theta)  \cos  (\delta\omega t+\delta\theta) \simeq 0$ under the integral. We can now straightforwardly calculate the FIM for the three parameters $\tau,\omega$ and $\theta$, invert the FIM and obtain the result given in the main text.

\section{Displaced squeezed state lidar FIM for homodyne detection}\label{app:sqFIM}
Now, let us come to the homodyne measurement and its FIM for the displaced squeezed state $\otimes_n \hat{D}_n \otimes_{n}\hat{S}_n\vert 0 \rangle$. Using the displacement vector $\vec{\mu}$ and the covariance matrix given in Eq.~(7) in the main text, we can calculate the FIM using the equation 
\begin{align}\label{app:FIMgaussian}
    F_{\alpha\beta} = \partial_{\alpha} \vec{\mu} \cdot \Sigma^{-1} \cdot \partial_{\beta} \vec{\mu} +  \frac{1}{2} \text{Tr}\left[ \Sigma^{-1}\cdot \partial_{\alpha} \Sigma \cdot\Sigma^{-1} \cdot \partial_{\beta} \Sigma \right]
\end{align}
for Gaussian distributions. Using the time bin basis, we can calculate this expression numerically, which we discuss in the later section \ref{app:numerical}. In this section, we seek analytical expressions. To do so, we change the basis from time bin to the mode basis, which allows us to analytically invert the covariance matrix. This inverted matrix can then be used to calculate the FIM using the more general formula for the FIM $F_{\alpha\beta}= \mathds{E} [\partial_{\alpha} \log (p(\vec{X}) ) \partial_{\beta} \log (p(\vec{X})) ]$ noting that $\log (p(\vec{X})) \sim -\frac{1}{2} (\vec{X}-\vec{\mu})^T \Sigma^{-1}(\vec{X} -\vec{\mu})$.

As we see in Eq.~\eqref{app:FIMgaussian}, we have two separate terms, the first containing only derivatives of the mean, while the second contains only terms of the covariance matrix. Because we do not have a mixed term, we can first assume $\vec{\mu}=0$ for notational convenience and  separately consider the case  $\vec{\mu} \neq 0$ later. 

Let us start. We  consider the zero mean case $ \langle \hat{X}(t) \rangle = 0$ in which the covariance function is given by $\Sigma (t,t') = \langle \hat{X}  (t) \hat{X}  (t') \rangle $. We will abbreviate $\lambda(t) =\omega_{\text{LO}}t+\theta_{\text{LO}}$.  Let us now calculate the covariance function. We begin with
\begin{align}
   2\langle \hat{X}(t) \hat{X} (t') \rangle = \Big\langle & \hat{E}(t) \hat{E}(t') e^{+i(\lambda (t)+\lambda(t'))} +\hat{E}^\dag(t) \hat{E}^\dag(t') e^{-i(\lambda (t)+\lambda(t'))} \\
   & + \hat{E} (t) \hat{E}^\dag(t') e^{i(\lambda (t)-\lambda (t'))} + \hat{E}^\dag (t) \hat{E} (t') e^{-i(\lambda (t)-\lambda (t'))} \Big\rangle ,
\end{align}
where the expectation values is with respect to the out state.
We will now use the out mode decomposition $\hat{E} (t) =\sum_n \phi_n (t) \hat{a}_n$ (omitting the out labels) and the transformation rules of squeezed states. We also remind the abbreviation $s_n =\sinh (r_n)$ and $c_n =\cosh (r_n)$. We find 
\begin{align}
    \langle \hat{E}(t) \hat{E}(t')\rangle = \sum_{n}\sum_m \phi_n (t) \phi_m (t') \langle 0 \vert \hat{S}^\dag \hat{a}_n \hat{a}_m \hat{S}\vert 0\rangle = - \sum_n \phi_n (t) \phi_n (t') c_n s_n e^{i \varphi_n}
\end{align}
\begin{align}
    \langle \hat{E}^\dag (t) \hat{E}^\dag (t')\rangle = \sum_{n}\sum_m \phi^*_n (t) \phi^*_m (t') \langle 0 \vert \hat{S}^\dag \hat{a}^\dag_n \hat{a}^\dag_m \hat{S}\vert 0\rangle = - \sum_n \phi^*_n (t) \phi^*_n (t') c_n s_n e^{-i \varphi_n}
\end{align}
\begin{align}
    \langle \hat{E} (t) \hat{E}^\dag (t') \rangle = \sum_{n}\sum_m \phi_n (t) \phi^*_m (t') \langle 0 \vert \hat{S}^\dag \hat{a}_n \hat{a}^\dag_m \hat{S}\vert 0\rangle =   \sum_n \phi_n (t) \phi^*_n (t') c_n^2 
\end{align}
\begin{align}
    \langle \hat{E}^\dag (t) \hat{E} (t') \rangle = \sum_{n}\sum_m \phi^*_n (t) \phi_m (t') \langle 0 \vert \hat{S}^\dag \hat{a}^\dag_n \hat{a}_m \hat{S}\vert 0\rangle =   \sum_n \phi^*_n (t) \phi_n (t') s_n^2  .
\end{align}
With the relation $c_n^2 - s_n^2 = 1$ and using the completeness relation $\sum_n \phi_n (t)\phi_n^*(t')= \delta (t-t')$ of our basis functions, we obtain the covariance function
\begin{align} \label{app:covfun}
    \langle \hat{X} (t) \hat{X} (t') \rangle = \frac{1}{2} \delta (t-t') + \sum_n s_n^2 \text{Re} \left[ \phi_n (t) \phi_n^* (t') e^{i (\lambda(t)-\lambda(t'))} \right] - \sum_n s_n c_n \text{Re} \left[ \phi_n (t) \phi_n(t') e^{i (\varphi_n + \lambda(t) + \lambda(t'))} \right] .
\end{align}
Let us now also discuss photon loss. We model the photon loss that may occur during the round trip due to diffraction, absorption or imperfect reflection with a beam splitter that is employed before measurement. We will then measure
\begin{align}
    \hat{E}' (t) = \sqrt{\kappa} \hat{E}(t) + \sqrt{1-\kappa} \hat{E}_B (t) ,
\end{align}
where $\kappa$ is the round trip transmissivity and $\hat{E}_B (t)$ is the environmental back ground mode with $\langle \hat{E}_B (t) \hat{E}_B^\dag (t') \rangle = \delta (t-t')$, while all other expectation values are zero. With this we can show that
\begin{align}
    \langle \hat{X}' (t) \rangle = \sqrt{\kappa}  \langle \hat{X} (t) \rangle
\end{align}
and 
\begin{align}
     \langle \hat{X}' (t) \hat{X}' (t') \rangle = \kappa \langle \hat{X} (t) \hat{X} (t') \rangle + \frac{1-\kappa}{2} \delta (t-t') .
\end{align}
Let us from now on omit the primes on the operators for notational convenience.
With this and Eq.\eqref{app:covfun}, we obtain the result given in the main text in Eq.~(7). We have assumed $\mu(t) = \langle \hat{X}(t) \rangle = 0$ in this calculation, but the same covariance matrix is found for $\mu(t) = \langle \hat{X}(t) \rangle \neq 0$.

By assuming finite time resolution $\Delta t$, and additionally assuming that $N_{\text{sq}}\bar{\phi}_n (t) \bar{\phi}_n (t') \cos (\delta\omega (t\pm t'))$ does not change a lot within our time window $\Delta t$, which implies $\delta\omega \Delta T \ll 1$,  we obtain the covariance matrix  
\begin{align}
    \langle \hat{X}_{i}\hat{X}_{j} \rangle  \simeq \frac{1}{2\Delta t} \delta_{i,j} + \sum_n \kappa s_n^2 \cdot \text{Re}\left[ \phi_{n,i}\phi_{n,j}^* e^{+i (\lambda_i-\lambda_j)} \right] -\sum_n  \kappa c_n s_n \cdot \text{Re} \left[  \phi_{n,i}  \phi_{n,j} e^{i (+\varphi_n +\lambda_i+\lambda_j)} \right] ,
\end{align}
where we introduced the discretized mode functions $\phi_{t,i}= \phi_{n} (i \Delta t)$ and $\lambda_i = \lambda (i \Delta t)$. We also introduce the discretized mode envelopes $\bar{\phi}_{t,i}= \bar{\phi}_{n} (i \Delta t)$ for upcoming calculations.
This covariance matrix can now be used to perform numerical calculations. However, we  also seek analytical results. To obtain these, we perform several steps of manipulation using the identities $\cos(x+y) = \cos (x)\cos (y)-\sin (x)\sin(y)$ and $\sin(x+y) = \sin (x)\cos(y) +\cos(x)\sin(y)$ and we finally arrive at
\begin{align}
     \langle \hat{X}_{i}\hat{X}_{ j} \rangle \simeq  \frac{1}{2\Delta t} \delta_{i,j} &+ \sum_n \bar{\phi}_{n,i} \cos (\delta\omega (t_i-\tau))  \bar{\phi}_{n,j} \cos (\delta\omega (t_j-\tau)) \left[ s_n^2 - c_n s_n  \cos (2\delta \theta + 2\delta\omega\tau+\varphi_n)\right] \\
     &+ \sum_n  \bar{\phi}_{n,i} \sin (\delta\omega (t_i-\tau))  \bar{\phi}_{n,j} \sin (\delta\omega (t_j-\tau)) \left[ s_n^2 + c_n s_n  \cos (2\delta \theta+ 2\delta\omega\tau +\varphi_n)\right] \\
     &+ \sum_n  \bar{\phi}_{n,i} \sin (\delta\omega (t_i-\tau))   \bar{\phi}_{n,j} \cos (\delta\omega (t_j-\tau))  \, c_n s_n  \sin (2\delta \theta+ 2\delta\omega\tau +\varphi_n) \\
        &+ \sum_n  \bar{\phi}_{n,i} \cos(\delta\omega (t_i-\tau))   \bar{\phi}_{n,j} \sin (\delta\omega (t_j-\tau))  \, c_n s_n  \sin (2\delta \theta+ 2\delta\omega\tau +\varphi_n) .
\end{align}  
To obtain analytical expressions, it is useful to invert this matrix. Before doing so, we make some approximations. First, we remind that we assumed $\delta\omega \Delta T \ll 1$, so that $\sin (\delta\omega (t-\tau)) \simeq \delta \omega (t-\tau)$ and $\cos (\delta \omega (t-\tau))\simeq 1$ for times around $\tau$. With this we find
\begin{align}
    \Bar{\phi}_{n,i} \sin (\delta \omega (t-\tau)) \simeq \delta \omega \cdot \Bar{\phi}_{n,i} \cdot (t_i -\tau) = \delta \omega \left[ \frac{1}{g} \sqrt{\frac{n}{2}}   \Bar{\phi}_{n-1,i}+\frac{1}{g} \sqrt{\frac{n+1}{2}}   \Bar{\phi}_{n+1,i}\right] ,
\end{align}
where we defined $g= \sigma/\sqrt{2}$ and used properties of our mode functions. With this, we find
\begin{align}
    \Sigma_{ij} = \langle \hat{X}_{i}\hat{X}_{j} \rangle \simeq  \frac{1}{2\Delta t} \delta_{i,j} &+ \sum_n \bar{\phi}_{n,i} \bar{\phi}_{n,j} \, A_n   \\
     &+ \delta\omega \sum_n \left( \alpha_n \bar{\phi}_{n-1,i} + \beta_n \bar{\phi}_{n+1,i}   \right) \, \bar{\phi}_{n,j} B_n   \\
     &+ \delta\omega \sum_n \bar{\phi}_{n,i}\left( \alpha_n \bar{\phi}_{n-1,j} + \beta_n \bar{\phi}_{n+1,j}    \right) B_n ,
\end{align}
with the new abbreviations $\alpha_n = \frac{1}{g} \sqrt{\frac{n}{2}}  $ and $\beta_n = \frac{1}{g} \sqrt{\frac{n+1}{2}}$, $A_n = \kappa ( s_n^2 - c_n s_n  \cos (2\delta \theta+2\delta\omega\tau +\varphi_n))$ and $B_n=  \kappa  c_n s_n  \sin (2\delta \theta +2\delta\omega\tau+\varphi_n)$. Now, let us note that the  discretized mode functions $\bar{\phi}_{n,i}$ are (approximate) unitary transformations $U$ on the discretized space and can be interpreted as coordinate transformations from time bin basis to the mode basis. We define $U_{n,i} := \sqrt{\Delta t} \bar{\phi}_{n,i}$ the (approximate) unitary coordinate transformation. The observed time bin data $X_i$ will then in the mode basis have components $Y_n = \sum_i U_{n,i} X_i \simeq \int \mathrm{d} t  \bar{\phi}_{n} (t) X(t)/\sqrt{\Delta t}$. Note, that the new basis depends on the parameter $\tau$ which will later become important. Now, we consider the transformation $U \Sigma U^\dag$, which, after some algebra, is found to be 
\begin{align}
     \tilde{\Sigma}_{nm}:= (U\Sigma U^\dag)_{nm} &=  \frac{1}{\Delta t} \Big(\delta_{nm} \left( A_n +1/2\right) \\
      & \quad +\delta_{n,m-1} \delta\omega \left( \alpha_m B_m +\beta_{m-1}B_{m-1} \right) \\
      & \quad +\delta_{n,m+1} \delta\omega \left( \alpha_{m+1} B_{m+1} +\beta_{m}B_{m} \right) \Big) ,
\end{align}
or as a matrix of which we only show the first few entries 
\begin{align}
 \tilde{\Sigma} =  \frac{1}{\Delta t} \left(
\begin{array}{cccc}
 A_0 +\frac{1}{2 } &  \delta \omega(\beta_0
    B_0 + \alpha_1 B_1 ) & 0 & 0 \\
 \delta \omega(\beta_0
    B_0 + \alpha_1 B_1 )  & A_1 +\frac{1}{2
    } & \delta \omega(\beta_1
    B_1 + \alpha_2 B_2 ) & 0 \\
 0 & \delta \omega(\beta_1
    B_1 + \alpha_2 B_2 ) & A_2  +\frac{1}{2
    } & \delta \omega(\beta_2
    B_2 + \alpha_3 B_3 ) \\
 0 & 0 & \delta \omega(\beta_2
    B_2 + \alpha_3 B_3 ) & A_3 +\frac{1}{2  } \\
\end{array}
\right)  .
\end{align} 
 Inverting this to first order in $\delta\omega /\sigma$, which corresponds to first order in $\delta\omega \Delta T$ due to $1/\sigma\sim \Delta T$, we obtain
\begin{align} \label{eq:invCov}
 \tilde{\Sigma}^{-1}= \Delta t \left(
\begin{array}{cccc}
 \frac{1}{A_0+\frac{1}{2}}  &
   -\frac{ \delta \omega (\beta_0  B_0 +\alpha_1 B_1)}{( A_0+1/2) ( A_1+1/2)} & 0 & 0 \\
 -\frac{ \delta \omega (\beta_0  B_0 +\alpha_1 B_1)}{( A_0+1/2) (A_1+1/2)} & \frac{1}{
   A_1+\frac{1}{2}}  & -\frac{ \delta \omega (\beta_1  B_1 +\alpha_2 B_2)}{(A_1+1/2) (A_2+1/2)} & 0 \\
 0 &-\frac{ \delta \omega (\beta_1  B_1 +\alpha_2 B_2)}{(A_1+1/2) (A_2+1/2)} &
   \frac{1}{A_2+\frac{1}{2}}  &
  -\frac{ \delta \omega (\beta_2  B_2 +\alpha_3 B_3)}{(A_2+1/2) (A_3+1/2)} \\
 0 & 0 &-\frac{ \delta \omega (\beta_2  B_2 +\alpha_3 B_3)}{(A_2+1/2) (A_3+1/2)} &  \frac{1}{A_3+\frac{1}{2}} \\
\end{array}
\right)  .
\end{align}
We will later see, that $\Delta t$ cancels out in the end result.
Now, we have the inverse covariance matrix in the mode basis. With this we can obtain the probability distribution of the transformed measurement outcomes $Y_n$, that are directly related with the time bin measurement outcome data $X_i$ via a unitary coordinate transformation. We have 
\begin{align}
    p(\vec{Y}) =   \frac{\exp ( -\frac{1}{2}\Vec{Y}^T \tilde{\Sigma}^{-1} \vec{Y})}{\left( (2\pi)^d \vert \tilde{\Sigma} \vert \right)^{1/2} } ,
\end{align}
where $d$ is the dimension of $Y$ and $\vert \tilde{\Sigma}\vert$ is the determinant of $\tilde{\Sigma}$. In principle, $d$ should be infinite, however, we assume here that it suffices to consider $d$ finite. We will later see, that indeed only the populated modes contribute to the FIM, which justifies us to neglect the rest.
The log likelihood $\log p (\Vec{Y})$ is needed to calculate the FIM
\begin{align}
     \log\left[ p (\vec{Y}) \right] &=    -\frac{\Delta t}{2} \left( \sum_n Y_n^2  \frac{1}{A_n + \frac{1}{2}} + \sum_n Y_{n}Y_{n+1} \frac{-2\cdot  \delta \omega (\beta_n  B_n +\alpha_{n+1} B_{n+1})}{( A_n+1/2) (A_{n+1}+1/2)}  \right) - \frac{1}{2} \log \left[ (2\pi)^d \vert \tilde{\Sigma} \vert\right] \\
     &= \sum_n Y_n^2  \frac{-\Delta t}{2A_n +1} + \sum_n Y_{n}Y_{n+1} \frac{  \Delta t \delta \omega (\beta_n  B_n +\alpha_{n+1} B_{n+1})}{(A_n+1/2) (A_{n+1}+1/2)}  - \frac{1}{2} \log \left[ (2\pi)^d \vert \tilde{\Sigma} \vert\right]\\
     &= \sum_n Y_n^2 D_n +\sum_n Y_n Y_{n+1} W_n  - \frac{1}{2}\log \left[ (2\pi)^d \vert \tilde{\Sigma} \vert\right]  ,
\end{align} 
where we introduced the new quantities $D_n$ and $W_n$. The FIM is given by $F_{ij} = \mathds{E}[ \partial_{\mu_i} \log p (Y)   \partial_{\mu_j} \log p (Y)]$. Let us calculate the derivatives and drop the terms proportional to $\delta \omega \Delta T$. Let us note that $\partial_{\mu_i} \log (\vert \Sigma \vert) = \text{Tr} [ \Sigma^{-1} \partial_{\mu_i} \Sigma]$ \cite{VanTrees2001}. With this, we obtain
\begin{align}
    \partial_\omega \log [ p(\vec{Y})] =  \sum_n Y_n^2 \partial_{\omega}D_n +\sum_n Y_n Y_{n+1} \partial_\omega W_n - \frac{1}{2}\text{Tr} \left[ \tilde{\Sigma}^{-1} \partial_\omega \tilde{\Sigma} \right] .
\end{align}
 For the time delay, we also have to take the derivative of $Y_n$, as our basis depends on $\tau$ (but not the other parameters), which yields 
 \begin{align}
     \partial_\tau Y_n =-\frac{\sigma}{2} (\sqrt{n} Y_{n-1} + \sqrt{n+1} Y_{n+1}) .
 \end{align}
 We thus have
 \begin{align}
       \partial_\tau \log [ p(\vec{Y})] =  \sum_n Y_n^2  \partial_\tau D_n + \sum_n  Y_n Y_{n+1}  \sigma \sqrt{n+1}\left( D_n - D_{n+1} \right) - \frac{1}{2} \text{Tr} \left[ \tilde{\Sigma}^{-1} \partial_\tau \tilde{\Sigma} \right]  .
 \end{align}
We also will calculate the derivative with respect to the phase. We find
\begin{align}
     \partial_\theta \log [ p(\vec{Y})] =  \sum_n Y_n^2  \partial_\theta D_n - \frac{1}{2} \text{Tr} \left[ \tilde{\Sigma}^{-1} \partial_\theta \tilde{\Sigma} \right]  .
\end{align}
Now, we have everything in order to do our analytical calculations for the FIM.
We first introduce again new notation to present results in a compact manner. We note that all derivatives of the log likelihood have the following structure:
\begin{align}
    \partial_{\mu_i} \log [ p(\vec{Y})] = \sum_n Y_n^2 G_{n,i} + \sum_n Y_{n} Y_{n+1} V_{n,i} - \frac{1}{2} \text{Tr} \left[ \tilde{\Sigma}^{-1} \partial_{\mu_i}\tilde{\Sigma} \right]  .
\end{align}
We thus obtain for the FIM for homodyne reception
\begin{align}
    F_{ij} &= \mathds{E}[ \partial_{\mu_i} \log p (Y)   \partial_{\mu_j} \log p (Y)] \\&= \sum_{n,m} \mathds{E}[ Y_n^2 Y_m^2] G_{n,i} G_{m,j} + \sum_n \mathds{E}[ Y_n^2 Y_{n+1}^2] V_{n,i} V_{n,j}  \\ &\quad -\frac{1}{2} \text{Tr} \left[ \tilde{\Sigma}^{-1} \partial_{\mu_i} \tilde{\Sigma} \right] \sum_n \mathds{E} \left[ Y_n^2 \right]  G_{n,j}    -\frac{1}{2} \text{Tr} \left[ \tilde{\Sigma}^{-1} \partial_{\mu_j} \tilde{\Sigma} \right] \sum_n \mathds{E} \left[ Y_n^2 \right]  G_{n,i} +  \frac{1}{4} \text{Tr} \left[ \tilde{\Sigma}^{-1} \partial_{\mu_i} \tilde{\Sigma} \right] \text{Tr} \left[ \tilde{\Sigma}^{-1} \partial_{\mu_j} \tilde{\Sigma} \right] ,
\end{align}
where terms such as $Y_{n}Y_{n+1}Y_{m}Y_{m+1}$ for $n\neq m$ and terms as $Y_{n}^2Y_{m}Y_{m+1}$ for all $n,m$ have expectation value $0$ and thus do not appear above. Furthermore, as we neglect terms of order $\delta \omega \Delta T$ or higher, $\tilde{\Sigma}_{nn}$ is diagonal and $\mathds{E}[Y_n Y_{n+1}] =0$. Let us note, that for Gaussian distributions the relation
\begin{align}
    \mathds{E} [Y_n Y_m Y_{n'} Y_{m'}] = \tilde{\Sigma}_{nm}\tilde{\Sigma}_{n'm'} + \tilde{\Sigma}_{nn'}\tilde{\Sigma}_{mm'} + \tilde{\Sigma}_{nm'}\tilde{\Sigma}_{mn'}
\end{align}
holds. Let us again note, that we drop in our final result terms of order $\delta\omega \Delta T$, which makes the covariance matrix diagonal, yielding 
\begin{align} \label{app:FIMp}
    F_{ij} &= \sum_{n=0}^\infty \tilde{\Sigma}_{nn} \tilde{\Sigma}_{n+1 n+1} V_{n,i} V_{n,j} +\sum_{n=0}^\infty 2 \tilde{\Sigma}_{nn}^2 G_{n,i} G_{n,j} + \left( \sum_{n=0}^\infty \tilde{\Sigma}_{nn} G_{n,i} \right) \left( \sum_{n=0}^\infty \tilde{\Sigma}_{nn} G_{n,j} \right) \\
    &\quad  - \frac{1}{2}\text{Tr} \left[ \tilde{\Sigma}^{-1} \partial_{\mu_i} \tilde{\Sigma} \right] \sum_{n=0}^\infty \tilde{\Sigma}_{nn} G_{n,j}   - \frac{1}{2} \text{Tr} \left[ \tilde{\Sigma}^{-1} \partial_{\mu_j} \tilde{\Sigma} \right] \sum_{n=0}^\infty \tilde{\Sigma}_{nn}  G_{n,i} + \frac{1}{4}  \text{Tr} \left[ \tilde{\Sigma}^{-1} \partial_{\mu_i} \tilde{\Sigma} \right] \text{Tr} \left[ \tilde{\Sigma}^{-1} \partial_{\mu_j} \tilde{\Sigma} \right]  ,
\end{align}
where $V_{n,\omega} = \partial_\omega W_n $, $G_{n,\omega} = \partial_\omega D_n$, $V_{n,\tau} =\sigma \sqrt{n+1} (D_n - D_{n+1})$, $G_{n,\tau} = \partial_\tau D_n$, $V_{n,\theta} = 0$ and $G_{n,\theta} = \partial_\theta D_n$. These coefficients can be found in the equations above. Let us now try to further simplify this expression. First, we note that neglecting terms of order $\delta\omega \Delta T$ yields
\begin{align}
    \text{Tr}\left[ \tilde{\Sigma}^{-1} \partial_{\mu_i} \tilde{\Sigma} \right] = \sum_{n=0}^\infty \tilde{\Sigma}_{nn}^{-1} \partial_{\mu_i} \tilde{\Sigma}_{nn} ,
\end{align}
as $\tilde{\Sigma}^{-1}$ is diagonal to this order. We also note, that $G_{n,i} = \partial_{\mu_i} D_n = -\frac{1}{2} \partial_{\mu_i} \tilde{\Sigma}_{nn}^{-1}$. Moreover, note that $ \text{Tr}\left[ \tilde{\Sigma}^{-1} \partial_{\mu_i} \tilde{\Sigma} \right] = -  \text{Tr}\left[ \tilde{\Sigma}\partial_{\mu_i} \tilde{\Sigma}^{-1}  \right]$. With this, we see that the last four terms in Eq.~\eqref{app:FIMp} are all proportional to each other and cancel each other out. We finally arrive at
\begin{align} \label{app:FIMsq}
    F_{ij} = \sum_{n=0}^\infty \tilde{\Sigma}_{nn} \tilde{\Sigma}_{n+1 n+1} V_{n,i} V_{n,j} + \frac{1}{2} \sum_{n=0}^\infty \tilde{\Sigma}_{nn}^2 \partial_{\mu_i} \tilde{\Sigma}^{-1}_{nn}   \partial_{\mu_j} \tilde{\Sigma}^{-1}_{nn}
\end{align}
and we have $V_{n,\omega} = - \partial_\omega \tilde{\Sigma}_{n, n+1}^{-1}$, $V_{n,\tau} = \frac{\sigma\sqrt{n+1}}{2} (\tilde{\Sigma}_{n+1,n+1}^{-1}-\tilde{\Sigma}_{n,n}^{-1})$ and $V_{n,\theta}=0$. Note, that all modes that are not populated, i.e. all modes with $r_n=0$, do not contribute to the FIM.
Our derivation has only assumed that $\delta\omega \Delta T \ll 1$ and is valid for arbitrary $\delta\theta$ and $\kappa$. The Heisenberg scaling in the case of $\kappa = 1$ comes  from terms such as 
\begin{equation}
   \frac{1}{\sinh^2 (r) - \cosh (r) \sinh(r) \cos(\delta\theta) + 1/2 } \overset{\delta\theta =0}{=} 2 e^{2r} \sim N_{\text{sq}}+1.
\end{equation}
Assuming $\delta\theta$ to be small such that $\cos (\delta\theta) \simeq 1 - \delta\theta^2/2$, we find the condition $e^{-2r} \gg \cosh(r)\sinh(r) \delta \theta^2$ to maintain Heisenberg scaling, which is equivalent to $\delta\theta \ll 1/(N_{\text{sq}}+1)$.

Let us now also discuss the general case with a general displaced squeezed state with an arbitrary collection of $\{\alpha_n \}_n$ and $\{ \zeta_n \}_n$. The mean is given by $\mu (t) = \sqrt{2\kappa} \text{Re}[s(t)e^{i(\omega_{\text{LO}}t+\theta_{\text{LO}})}] =\sqrt{2\kappa}  \text{Re}[ \sum_n \alpha_n \Bar{\phi}_n^{\text{out}}(t) e^{i (\delta \omega t+\delta \theta)}]$ using the fact that the returned signal can be decomposed as $s(t) = \sum_n \alpha_n \phi_n^{\text{out}} (t)$. With finite time resolution $\Delta t$, and assuming that $s(t)e^{i(\omega_{\text{LO}}t+\theta_{\text{LO}})}$ does not vary much within that time window, we obtain $\mu_i = \langle \hat{X}_i\rangle \simeq  \sqrt{2\kappa} \text{Re}[s(t_i)e^{i(\omega_{\text{LO}}t_i+\theta_{\text{LO}})}] =\sqrt{2\kappa} $ with $t_i = i\Delta t$.
The term contributed by displacement appearing in the FIM is $\partial_\alpha \vec{\mu}^T \cdot \Sigma^{-1} \cdot \partial_\beta \vec{\mu}$. We seek analytical expressions, which may be difficult to obtain using the time bin basis. So let us perform a change of basis, using our earlier defined $U_{n,i} = \sqrt{\Delta t} \bar{\phi}_n (t_i)$. Let us note, that we have
 \begin{align}
     \partial_\alpha \vec{\mu}^T \cdot \Sigma^{-1} \cdot \partial_\beta \vec{\mu} = \partial_\alpha \vec{\mu}^T U^\dag \underbrace{U \Sigma^{-1} U^\dag}_{\tilde{\Sigma}^{-1}} U \partial_\beta \Vec{\mu} ,
 \end{align}
where we already calculated $\tilde{\Sigma}^{-1}$ earlier. So the only remaining term to calculate is $U \partial_\beta \Vec{\mu}$. 
For a general signal $s(t)$, we have
\begin{align}
      \left( U \cdot \partial_\alpha \vec{\mu} \right)_n &= \frac{\sqrt{2}}{\sqrt{\Delta t}}\sum_i \Delta t \Bar{\phi}_n (t_i) \partial_\alpha \text{Re} \left[ s(t_i)  e^{i(\omega_{\text{LO}} t_i +\theta_{\text{LO}})} \right] \\
      &\simeq \frac{\sqrt{2}}{\sqrt{\Delta t}}\int \mathrm{d}t\, \Bar{\phi}_n (t) \partial_\alpha \text{Re} \left[ s(t)  e^{i(\omega_{\text{LO}} t+\theta_{\text{LO}})} \right] ,
\end{align}
reminding $t_i = i \Delta t$.
With this, and $\tilde{\Sigma}^{-1}$ from Eq.~\eqref{eq:invCov}, we obtain the FIM of a general displaced squeezed state
\begin{align}
    F_{\alpha\beta} = \left( U\cdot \partial_\alpha \vec{\mu} \right)^T \tilde{\Sigma}^{-1} U\cdot \partial_\beta \vec{\mu} + \sum_{n=0}^\infty \tilde{\Sigma}_{nn} \tilde{\Sigma}_{n+1 n+1} V_{n,\alpha} V_{n,\beta} + \frac{1}{2} \sum_{n=0}^\infty \tilde{\Sigma}_{nn}^2 \partial_{\alpha} \tilde{\Sigma}^{-1}_{nn}   \partial_{\beta} \tilde{\Sigma}^{-1}_{nn} ,
\end{align}
with $V_{n,\omega} = - \partial_\omega \tilde{\Sigma}_{n, n+1}^{-1}$, $V_{n,\tau} = \frac{\sigma\sqrt{n+1}}{2} (\tilde{\Sigma}_{n+1,n+1}^{-1}-\tilde{\Sigma}_{n,n}^{-1})$ and $V_{n,\theta}=0$. The expression for the FIM is valid for $\delta\omega \Delta T\ll 1$, arbitrary $\delta\theta$ and arbitrary $\kappa$.

Let us also consider the special case of displacement signals with no frequency modulation, i.e. $\mu_i = \sqrt{2} \bar{s}(t_i) \cos (\delta\omega_{\text{coh}}(t-\tau)+\delta\omega_{\text{coh}}\tau+\delta \theta_{\text{coh}})$, where $\bar{s}(t)$ is the envelope of the signal and we defined $\delta\omega_{\text{coh}}= \omega_{\text{LO}}- \omega_{\text{coh}}$ and $\delta\theta_{\text{coh}}= \theta_{\text{LO}}- \theta_{\text{coh}}$, where $\omega_{\text{coh}}$ and $\theta_{\text{coh}}$ are the carrier frequency and the phase of our displacement operation. Then, we find
\begin{align}
    \left( U \cdot \partial_\alpha \vec{\mu} \right)_n &= \frac{\sqrt{2}}{\sqrt{\Delta t}} \sum_i \Delta t \Bar{\phi}_n (t_i) \partial_\alpha \left[ \bar{s}(t_i) \cos (\delta\omega_{\text{coh}}(t_i-\tau)+\delta\omega_{\text{coh}}\tau+\delta \theta_{\text{coh}}) \right] \\
    &\simeq \frac{\sqrt{2}}{\sqrt{\Delta t}}\int \mathrm{d}t\, \Bar{\phi}_n (t) \partial_\alpha \left[ \bar{s}(t) \cos (\delta\omega_{\text{coh}}(t-\tau)+\delta\omega_{\text{coh}}\tau+\delta \theta_{\text{coh}}) \right] .
\end{align}

Let us now calculate the FIM for the  specific displaced squeezed state considered in the main text, i.e. $\hat{D}_1 (\sqrt{N_{\text{coh}}}e^{-i\pi/4}) \hat{S}_{0}(r)\hat{S}_1 (r e^{-i\pi/2}) \hat{S}_2 (r) \vert 0 \rangle$. We have  $s(t) = \sqrt{N_{\text{coh}}}\phi_1(t) e^{-i\pi /4}$.  The total photon number is $N = N_{\text{coh}} + N_{\text{sq}}$ with $N_{\text{sq}} = \sum_n \sinh^2 (r_n) = 3\sinh^2 (r)$. We also have shown before that $\Delta T = 3/\sigma^2$ and $\Delta\omega^2 = 3\sigma^2 /4$. In the case of $\delta\omega \Delta T\ll 1 $ and $ \vert\delta\theta +\delta\omega \tau \vert\ll 1$, we obtain
\begin{align}
     \left( U \cdot \partial_\tau \vec{\mu} \right)_n &= \sqrt{N_{\text{coh}}} \left[ \gamma_{\tau 1 0} \delta_{n0} + \gamma_{\tau 1 2} \delta_{n2}\right] \\
     \left( U \cdot \partial_\omega \vec{\mu} \right)_n & =  \sqrt{N_{\text{coh}}}\frac{1}{-i} \left[\gamma_{\omega 1 0} \delta_{n0} + \gamma_{\omega 1 2} \delta_{n2} + \gamma_{\omega 1 1} \delta_{n1} \right] \\
     \left( U \cdot \partial_\theta \vec{\mu} \right)_n &= \sqrt{N_{\text{coh}}} \delta_{n1} ,
\end{align}
where the $\gamma_{\alpha nk}$ are given in section~\ref{app:Modes}.
We can abbreviate this as $\left( U \cdot \partial_\alpha \vec{\mu} \right)_n = \bar{\gamma}_{\alpha 10 } \delta_{n0}+\bar{\gamma}_{\alpha 11 } \delta_{n1}+ \bar{\gamma}_{\alpha 12} \delta_{n2}$. Dropping again terms of order $\delta\omega \Delta T$, which means that $\tilde{\Sigma}$ is diagonal, we obtain
\begin{align}\label{app:FIMdisp1}
    \left( U\cdot \partial_\alpha \vec{\mu} \right)^T \tilde{\Sigma}^{-1} U\cdot \partial_\beta \vec{\mu} =  \sum_{n=0}^2 \bar{\gamma}_{\alpha 1 n} \bar{\gamma}_{\beta 1 n} \frac{1}{A_n + 1/2} .
\end{align}
One can straightfowardly show that the diagonal terms have Heisenberg like scaling $\sim N_{\text{coh}}\cdot N_{\text{sq}}$, when the LO's frequency and phase are matched to the returned state and when $\kappa =1$. By now adding the above result with the squeezing term given in Eq.~\eqref{app:FIMsq}, we obtain the FIM for our state $\hat{D}_1 (\sqrt{N_{\text{coh}}}e^{-i\pi/4}) \hat{S}_{0}(r)\hat{S}_1 (r e^{i\pi/2}) \hat{S}_2 (r) \vert 0 \rangle$. The expression for the FIM is valid for $\delta\omega \Delta T\ll 1 $ and $ \vert\delta\theta +\delta\omega \tau \vert\ll 1$ with  arbitrary $\kappa$.

\section{Displacement dominate case} \label{app:displacementdominant}
Let us now further discuss the displacement dominant case $N_{\text{coh}}\gg N_{\text{sq}}$ of our QL, where most of the photons come from the displacement. Then, all the dominant contributions of the FIM will come from the displacement term and the other terms from the squeezing can be neglected. We then invert the FIM given in Eq.~\eqref{app:FIMdisp1}, assuming $\vert \delta\theta +\delta\omega \tau\vert \ll 1/(N_{\text{sq}}+1)$ for $\kappa =1$  (this condition is needed to preserve the enhancment contained in the $A_n$ coefficients), and obtain
\begin{align}\label{eq:QCRBdis}
    F^{-1}_{\hphantom{-1}\tau\tau} = \frac{9}{16} \frac{1-\kappa + \kappa e ^{-2r}}{\Delta\omega^2 \kappa N_{\text{coh}}} \quad\quad\quad\quad   F^{-1}_{\hphantom{-1}\omega\omega} =\frac{9}{16} \frac{1-\kappa +\kappa e ^{-2r}}{\Delta T^2 \kappa N_{\text{coh}}}  ,
\end{align}
where $e^{2r}= \frac{2}{3}N_{\text{sq}}+1+2\sqrt{\frac{N_{\text{sq}}}{3}(1+\frac{N_{\text{sq}}}{3})}$. The prefactor $9/16$ is approximately equal to the prefactor $1/2$ of the CL when heterodyne detection is employed. In the ideal case $\kappa = 1$, we get $ F^{-1}_{\hphantom{-1}\tau\tau} \simeq 1/ 2\Delta\omega^2 N_{\text{coh}} e^{2r}$ and $F^{-1}_{\hphantom{-1}\omega\omega} \simeq 1/ 2\Delta T^2 N_{\text{coh}}e^{2r}$. We thus have an enhancing factor of $e^{-2r}$ in the CRBs.  In the non-ideal case with photon loss and the condition $(1-\kappa)/\kappa \gg e^{-2r}$ (which is basically satisfied for $r\simeq 1$ and $\kappa <0.9$), the constraints on the detunings are looser  $\vert \delta\theta +\delta\omega \tau\vert \ll 1$, which is the same constraint a classical protocol with homodyne has to satisfy. Thus, we do have the same requirements for tuning the homodyne receiver for quantum and classical protocol, yet we have a quantum advantage of $F^{-1}_{\hphantom{-1}\tau\tau} \simeq (1-\kappa)/ 2\Delta\omega^2 \kappa N_{\text{coh}}$ and $F^{-1}_{\hphantom{-1}\omega\omega} \simeq (1-\kappa)/ 2\Delta T^2 \kappa N_{\text{coh}}$. In this case, we get an enhancing factor of $1-\kappa$ for the CRBs compared to the CL heterodyne strategy and we only need squeezing of about $r\simeq 1$ (which is experimentally achievable \cite{presutti}) to achieve the full quantum advantage.

 \section{Estimators}\label{app:estimators}
So far we have only discussed the optimal measurement. Here we discuss the estimators that can be used to map the measurement data to estimates of the parameters. For this, we assume to have prior estimates $\tau^{\text{pr}}$, $\omega^{\text{pr}}$ and $\theta^{\text{pr}}$. These priors could for example be obtained by first using a classical coherent state with heterodyne detection. The respective MSEs for these priors would then be give by Eq.~(9) of the main text. Now, we assume the priors to be good enough such that we can tune the homodyne receiver so that $\vert\delta\omega \Delta T \vert \ll 1/\sqrt{N_{\text{sq}}+1}$ and  $\vert\delta\theta +\delta\omega\tau \vert \ll 1/(N_{\text{sq}}+1)$, where $\delta\theta = \theta^{\text{LO}}-\theta^{\text{pr}}$ and $\delta\omega=\omega^{\text{LO}}-\omega^{\text{pr}}$ and we further require $\vert \tau -\tau^{\text{pr}}\vert \Delta\omega \ll 1/(N_{\text{sq}}+1)$. 
The raw measurement data we receive when we measure the quantum state with the homodyne receiver tuned with the priors is then given by $X(t)$. From this, we construct by classical post processing $x_n^{\text{pr}} := \int \mathrm{d}t\, X(t) \Phi_n (t;\tau^{\text{pr}},\omega^{\text{pr}},\theta^{\text{pr}})e^{i(\omega^{\text{pr}} t+\theta^{\text{pr}})}$. This classical variable is associated with the quantum operator $\hat{x}_n^{\text{pr}}=(\hat{a}_n^{\text{pr}}+(\hat{a}_n^{\text{pr}})^\dag)/\sqrt{2}$, where $\hat{a}_n^{\text{pr}} = \int\mathrm{d} t \Phi_n^*(t;\tau^{\text{pr}},\omega^{\text{pr}},\theta^{\text{pr}}) \hat{E}(t)$. Using the inverse of Eq.~(5) of the main text $\hat{a}_n^{\text{pr}} \approx A_n^* \hat{a}_n^{\text{out}} - B_n \hat{a}_{n-1}^{\text{out}} +B_{n+1}^* \hat{a}_{n+1}^{\text{out}}$  (which is valid due to the assumed quality of our priors)  and the out state $\vert \psi^{\text{out}}\rangle$, we can easily calculate the moments of the random variables $\{ x_n^{\text{pr}}\}_n$, for example the first moments $\mathds{E}[x_n^{\text{pr}}]=\langle \psi^{\text{out}}\vert \hat{x}_n^{\text{pr}} \vert \psi^{\text{out}}\rangle$. We found that $\tau_{\text{est}} =   \tau^{\text{pr}}-(\frac{1}{\sqrt{2}}x_2^{\text{pr}}-x_0^{\text{pr}})/(\sqrt{\kappa N_{\text{coh}}}\sigma)$ and $\omega_{\text{est}} = \omega^{\text{pr}}-(x_0^{\text{pr}} +\frac{1}{\sqrt{2}}x_2^{\text{pr}})\sigma/(2\sqrt{\kappa N_{\text{coh}}}) $ are approximately unbiased estimators. Let us interpret this result. Our mode transform given in Eq.~(5) of the main text can be written as $\hat{a}_n^{\text{out}} \approx \hat{a}_n^{\text{pr}} + \epsilon_\tau \partial_\tau \hat{a}_n^{\text{pr}} + \epsilon_\omega \partial_\omega \hat{a}_n^{\text{pr}} + \epsilon_\theta \partial_\theta \hat{a}_n^{\text{pr}}$ (the "in" label got exchanged with the "pr" label, showing that estimation with good prior knowledge is equivalent to estimating small deviations). Now, the derivative modes are modes in their own right after normalization. We find that the  displacement in $\hat{a}_1^{\text{out}}$ gets transferred to the modes $\hat{a}_0^{(\text{pr})} \pm \frac{1}{\sqrt{2}} \hat{a}_1^{(\text{pr})}$, which are the derivative modes of the mode $\hat{a}_1^{(\text{pr})}$ with respect to time delay and frequency shift, i.e. $\partial_\tau \hat{a}_1^{(\text{pr})} \sim \hat{a}_0^{(\text{pr})} - \frac{1}{\sqrt{2}} \hat{a}_1^{(\text{pr})}$ and $\partial_\omega \hat{a}_1^{(\text{pr})} \sim \hat{a}_0^{(\text{pr})} + \frac{1}{\sqrt{2}} \hat{a}_1^{(\text{pr})}$. The real part of the displacement (i.e. the expectation value of the position quadrature) of mode $\partial_\tau \hat{a}_1^{(\text{pr})}$ is proportional to the value of $\tau-\tau^{\text{pr}}$ and the real part of the displacement of mode $\partial_\omega\hat{a}_1^{(\text{pr})}$ is proportional to $\omega -\omega^{\text{pr}}$, the parameters we want to estimate. This explains our choice of estimators. A quantum enhancement is to be expected if the corresponding position quadratures of the derivative modes are squeezed below the vacuum level, increasing the signal to noise ratio compared to the classical case. With our particular mode configuration, this is indeed the case. It is straightforward to calculate the variance of the estimators  $\text{Var}[\tau_{\text{est}}]\approx 2/\Delta\omega^2 N^2$ and $\text{Var}[\omega_{\text{est}}]\approx 2/\Delta T^2 N^2$ for $N_{\text{sq}}= 3N/4$ and $\kappa =1$. We thus reach the HL simultaneously in a single shot. In the case $\kappa \lesssim 0.9$ and $N\gg 1$, we find that it is most beneficial to allocate most of the energy into the displacement $N_{\text{coh}}\gg N_{\text{sq}}$ and a squeezing of $r\simeq 1$ is enough to obtain full quantum advantage. We find that the above given estimators are optimal in this scenario, and the requirements for the detunings become less restrictive $\vert\delta\theta +\delta\omega\tau \vert \ll 1$, which is the same requirement as that of a completely classical protocol that employs homodyne detection.

 \section{Numerical calculations and heterodyne detection for the QL} \label{app:numerical}
Let us now come to perform numerical calculations for our quantum lidar.  The measurement data $X(t)$ is obtained from measuring $\hat{X}(t)$.    By assuming that our measurement device averages over time windows $\Delta t$, we obtain operators $\hat{X}_i = \int_{i\Delta t}^{(i+1)\Delta t}\mathrm{d}t\, \hat{X}(t)/\Delta t$, and the corresponding measurement data $X_i$ is now discrete. As we have shown previously in Appendix~\ref{app:homodyne}, the data is Gaussian distributed with mean $\mu_i=\langle\hat{X}_i \rangle$ and covariance matrix $\Sigma_{ij}=\langle\hat{X}_i \hat{X}_j \rangle$.  In principle $i\in \mathds{Z}$, but for our numerical calculations, we have to select a finite number of time bins $i\in [ i_{\text{min}}, i_{\text{max}} ]$ and thus a finite measurement duration $  (i_{\text{max}}-i_{\text{min}})\Delta t$. In our numerical calculations, we assume that the time duration $\Delta T$ of the signal is much smaller than the measurement time, that is $\Delta T \ll (i_{\text{max}}-i_{\text{min}})\Delta t$, and that the signal is fully contained within our entire measurement window. We introduce the notation $t_i =i \Delta t $. With this, we find   $\mu_i =\langle \hat{X}_i\rangle$ and the covariance matrix  can be approximated as
\begin{align}
   \Sigma_{ij} =  \langle \hat{X}_i \hat{X}_j\rangle  \simeq \frac{1}{2\Delta t} \delta_{ij}+ \kappa\sum_{n=0}^\infty \bar{\phi}_n (t_i)\bar{\phi}_n (t_j) \left[ 
N_n \cos \left( \delta\omega \, (t_i-t_j) \right) - \sqrt{N_n (N_n+1)} \cos \left( \delta\omega \, (t_i+t_j)+2\delta\theta+\varphi_n \right) \right] ,
\end{align}
assuming that the function  $\Bar{\phi}_{n} (t_i) \cos (\delta\omega (t_i \pm t_j))$ does not vary much with in the time window $\Delta t$.
Thus, the Gaussian probability distribution of measurement outcomes $X_i$, which we collect in the vector $\vec{X}$, is 
\begin{align}
     p(\vec{X}) \sim \exp \left(-\frac{1}{2} (\vec{X}-\vec{\mu} )^T\Sigma^{-1} (\Vec{X}-\vec{\mu} )\right) .
\end{align}
In our main text, we have considered the state $\hat{D}_1 (\sqrt{N_{\text{coh}}}e^{-i\pi/4}) \hat{S}_{0}(r)\hat{S}_1 (r e^{-i\pi/2}) \hat{S}_2 (r) \vert 0 \rangle$. In our numerical calculations, we assumed a measurement of $100$ time bins with $\Delta t = 0.2$ (arbitrary units) with the parameters chosen to be $\tau =0$, $\omega =5$, $\sigma =1$. With this, the time duration of the pulse is $\Delta T = \sqrt{3}$ (arbitrary units) and measurement time $20$ (arbitrary units).
We used this framework  to numerically calculate the FIM for the parameters $\tau,\omega,\theta$. We simply use the mean and covariance matrix $\Sigma_{ij}$ to calculate the FIM with $ F_{\alpha \beta} = \partial_{\alpha} \vec{\mu} \cdot \Sigma^{-1} \cdot \partial_{\beta} \vec{\mu} +  \frac{1}{2} \text{Tr}\left[ \Sigma^{-1}\cdot \partial_\alpha \Sigma \cdot\Sigma^{-1} \cdot \partial_\beta \Sigma \right]$. We then inverted this matrix to obtain $F^{-1}_{\hphantom{-1} \tau\tau}$ and $F^{-1}_{\hphantom{-1} \omega\omega}$ from which we calculate $\text{Var}[\tau]\text{Var}[\omega]\simeq F^{-1}_{\hphantom{-1} \tau\tau} F^{-1}_{\hphantom{-1} \omega\omega}$.  
All analytical results were successfully reproduced using this numerical procedure.

We also studied heterodyne reception for our QL, i.e. $\delta\omega\Delta T \gg 1$ with arbitrary phase detuning $\delta\theta$ with $N_{\text{sq}}=N$ and $N_{\text{coh}}=0$. Our simulations have revealed that in this setup, both MSEs scale according to the standard quantum limit, i.e. $ \text{Var}[\tau]\text{Var}[\omega]\sim N^{-2}$. For the state considered in the main text, we find that QL with heterodyne performs similar to the CL with heterodyne, with no quantum advantage for the joint estimation task.
However, by changing the squeezing angle of mode $1$ to $\varphi_0 = 0$ and setting $N_{\text{sq}}=N$ we found that the QL with heterodyne outperforms the CL heterodyne strategy for frequency estimation with a constant factor advantage $F^{-1}_{\hphantom{-1} \omega\omega} = 1/ 3 \Delta\omega^2 N$, but performs worse for time-delay estimation.  Thus, with appropriate mode configuration, we have found a global estimation strategy (one that does not require the tuning of $\omega_{\text{LO}}$ and $\theta_{\text{LO}}$) that outperforms the best-known classical global heterodyne protocol for single parameter estimation. With $\kappa\ll 1$, the performances of QL with heterodyne and CL with heterodyne are approximately equal.

\end{widetext}

\end{document}